 \documentclass[12pt,a4paper]{article}
\input epsf
\usepackage{amsmath}
\usepackage{amssymb}
\usepackage{graphicx}
\def\IC{\mathbb{C}   }

\def\CC{{\cal C}}
\def\CD{{\cal D}}
     \def\e{\epsilon}
\def\uu{ u}
 \def\hf{\frac{1}{2}}
\def\({\left(}
\def\){\right)}
\def\<{\left\langle\,}
\def\>{\, \right\rangle}
\def\[{\left[}
\def\]{\right]}
\def\xpm{ x^{\pm} }
\def\xp{ x^{+} }
\def\xm{ x^{-} }
  
  \def\sr{  \sigma}
  \def\Li{{\rm Li}}
  \def\a{\alpha}\def\aa{\a^{-1} }

\newcommand{\eqal}[2]{\begin{eqnarray} #2 \end{eqnarray}}

\newcommand{\eqn}[2]{\begin{equation} #2 \end{equation}}

\newcommand{\no}{\nonumber}

\def\la{\label}
\def\sgn{\;{\rm sign}\:}

\def\sn{{\rm sn\,} }

  \def\cn{{\rm cn\,} }
  \def\dn{{\rm dn\,} }

     \def\sd{{\rm sd\,} }
    \def\cd{{\rm cd\,} }
     \def\cs{{\rm cs\,} }
     \def\nd{ {\rm nd\,}}
       \def\ds{{\rm ds\,} }
        
           \def\ns{{\rm ns\,} }
\def\K{ {\rm K}}
\def\o{\omega}
\newcommand{\CK}{{\cal K}}
\def\be{\begin{eqnarray}}
\def\ee{\end{eqnarray}}
\def\bi{\begin{itemize}}
\def\ei{\end{itemize}}

\def\Kmm{ K^{_{<<}}}
\def\Kpm{ K^{_{><}}}
\def\Kmp{K^{_{<>}}}
\def\Kpp{ K^{_{>>}}}
\def\rhp{ \rho^{_{>}}}
\def\rhm{\rho^{_{<}}}

\def\tmm{ \theta^{_{<<}}}
\def\tpm{ \theta^{_{><}}}
\def\tmp{\theta^{_{<>}}}
\def\tpp{ \theta^{_{>>}}}

\def\vppm{ \varphi^{_{><}}}
\def\vpmp{\varphi^{_{<>}}}
\def\vppp{ \varphi^{_{>>}}}
\def\CKmm{ \CK^{_{<<}}}

\def\CKmp{\CK^{_{<>}}}
\def\CKpp{ \CK^{_{>>}}}

\newcommand{\pint}{\makebox[0pt][l]{\hspace{3.1pt}$-$}\int}

\newcommand{\sfrac}[2]{{\textstyle\frac{#1}{#2}}}
\newcommand{\half}{\sfrac{1}{2}}
\newcommand{\Tr}{{\rm Tr \,}}

  \def\vp{\varphi}
 \def\p{\partial}
 \def\s{  s }

    \def \xp{ x^+ }
     \def \xm{ x^-}
     \def \xpm{ x^\pm}
     \def\th{\theta}
     \def\CO{{\cal O}}

\def\b{\beta}

\def\y{ y} 
 
 \def\srb{\bar \sr}    %= \s-{1\over 2\pi g}}

\begin{document}
\thispagestyle{empty}
\begin{flushright}
{hep-th/0703031} \\
{SPhT-T07/029}\\
{  NSF-KITP-07-29}
\end{flushright}
\vspace{1cm}
\setcounter{footnote}{0}
\begin{center}
{\Large{\bf Strong coupling limit of Bethe Ansatz equations
}}

\vspace{20mm}
{\sc  Ivan Kostov$^{\ast \bullet \star }$, 
Didina Serban$^{\ast\bullet}$ 
and Dmytro Volin$^{\ast \circ}$} 
\\[7mm]
$^\ast ${\it Service de Physique Th\'eorique, CNRS-URA 2306
\\
C.E.A.-Saclay \\
F-91191 Gif-sur-Yvette, France}\\[1mm]
{\it and} \\ [1mm]
$^\bullet ${\it Kavli Institute for Theoretical Physics,
University of California\\
Santa Barbara, CA 93106 USA} 
\\[1mm]
{\it and} \\ [1mm]
$^\circ${\it Bogolyubov Institute for Theoretical Physics\\
 14b Metrolohichna Str. \\
Kyiv, 03143 Ukraine} \\
[10mm]
{\sc Abstract}\\[2mm]
\end{center}

\noindent{ We develop a method to analyze the strong coupling limit of
the Bethe ansatz equations supposed to give the spectrum of anomalous
dimensions of the planar ${\cal N}=4$ gauge theory.  This method is
particularly adapted for the three rank-one sectors, $su(2)$,
$su(1|1)$ and $sl(2)$.  We use the elliptic parametrization of the
Bethe ansatz variables, which degenerates to a hyperbolic one in the
strong coupling limit.  We analyze the equations for the highest
excited states in the $su(2)$ and $su(1|1)$ sectors and for the state
corresponding to the twist-two operator in the $sl(2)$ sector, both
without and with the dressing kernel.  In some cases we were able to
give analytic expressions for the leading order magnon densities.  Our
method reproduces all existing analytical and numerical results for
these states at the leading order.  }
 
\vfill

{\small $^\star$Associate member of the {\it Institute for Nuclear
Research and Nuclear Energy, Bulgarian Academy of Sciences, 72
Tsarigradsko Chauss\'ee, 1784 Sofia, Bulgaria}}

\newpage

\setcounter{page}{1}

% \tableofcontents
 
 \section{Introduction}
\label{sec:intro}

There is mounting evidence that the planar ${\cal N}=4$ SYM gauge
theory is integrable \cite{MZ02,BS03,BKS} to higher orders in
perturbation theory, nourishing the hope to prove by this mean its
equivalence with the string theory on $AdS_5\times S^5$
\cite{MaldaAdS,GKP98,Witten98}.  Under the assumption of
integrability, the spectrum of the anomalous dimensions of the planar
${\cal N}=4$ SYM gauge theory is encoded, at least for states with
large number of operators, in the Bethe ansatz equations.  The
$S$-matrix behind these equations \cite{StaudS} is fixed by symmetry
\cite{BSansaetze, Beisert05}, up to a scalar factor which is supposed
to obey a crossing symmetry relation \cite{Janik}.  At strong
coupling, the scalar factor should allow to reproduce the string
theory results.  Its first two orders in the strong coupling expansion
were fixed in this way \cite{AFS,HL}.  More recently, based on
previous work of Beisert, Hern\'andez and L\'opez \cite{BHL}, Beisert,
Eden and Staudacher \cite{BES} proposed an expression for the scalar
factor which is crossing symmetric and reproduces both the strong
coupling, string results and the perturbative results for the
anomalous dimensions in the gauge theory.  An efficient way of testing
the predictions of the Bethe ansatz equations proved to be via the
anomalous dimension of the twist-two operator.  In the regime of large
Lorentz spin $S$, this quantity scales logarithmically
\begin{equation}
\Delta-S=f(g) \ln S +\ldots\;.
\end{equation}
The universal scaling function for ${\cal N}=4$ SYM up to third loop
order was extracted \cite{KLOV} from a perturbative three-loop
computation in QCD \cite{Moch}.  The same quantity appears in the
iterative structure \cite{Bern03} of the multi-gluon amplitudes of the
supersymmetric gauge theory.  It was computed up to three loop order
in \cite{Bern05}, and, by an impressing effort, up to four loop order
in \cite{Bern06,CSV}.  The four loop result agrees with the prediction
based on the proposal \cite{BES}.  On the other hand, at strong
coupling the universal scaling function $f(g)$ is predicted to behave
as \cite{GKP,FT}
 \begin{equation}
 \la{fdeg}
f(g)=4g -\frac{3\ln2}{\pi}+\ldots \;.
\end{equation}
Here we adopt the recent convention for the coupling constant $g$
\begin{equation}
g^2=\frac{\lambda}{16\pi^2}\;.
\end{equation}
After the proposal \cite{BES} was made, several groups \cite{KL06,
Benna-I, Benna2} attempted to derive from it the strong coupling
expression (\ref{fdeg}).  While the numerical work \cite{Benna-I}
easily reproduced the behavior (\ref{fdeg}) and even predicted the
next term in the expansion, the analytical treatment proved to be much
more difficult.  Up to now, only the first term was obtained
analytically \cite{KL06, Benna2}.

One of the aspects which deserve further attention is to understand
the origin of the dressing phase \cite{BES} and to test its validity.
It is believed that the dressing phase comes from a non-trivial
structure of the vacuum.  Very recently, a structure similar to that
of the dressing phase was obtained \cite{RSZ}, via the nested Bethe
ansatz, for one of the non-trivial ``vacuum'' states in one of the
sectors which are not of rank one.

The aim of the present paper is to develop a systematic approach for
solving the Bethe ansatz equations at strong coupling.  Our analysis
is based on the formulation of the Bethe equations as integral
equations, and it works most simply on, although it is not restricted
to, states as the antiferromagnetic state in the $su(2)$ sector, on
the most excited state in the $su(1|1)$ sector, or for the
finite-twist operators, in the $sl(2)$ sector.  Some of this cases
were already studied numerically or analytically; here we aim to a
unitary treatment and hope that this method will be useful to obtain
systematical $1/g$ expansion.

Our starting point is the elliptic parametrization of the variables
$u$ and $\ x^\pm$ that appear in the Bethe equations.  The elliptic
parametrization we use here is related to the one proposed by Janik
\cite{Janik} by a Gauss-Landen transformation, and it appeared
independently in \cite{Beisert05} and \cite{Gomez}.  The elliptic
modulus is defined by
\begin{equation}
\frac{k'}{k}=\frac{1}{4g}\equiv \e\;.
\end{equation}
In the hyperbolic  limit $g\to \infty$, this parametrization   supplies  natural variables in the different regimes,  recently discussed
in \cite{BHL,Maldacena-Swanson}.

As it became clear from the numerical solutions of the integral
equations \cite{ES}, the points $u=\pm 2g$ become singular for large
$g$ and they split the real axis $u$ into two regions, with $|u|/2g$
less or larger than one, where two different hyperbolic
parametrizations apply.\footnote{The two hyperbolic parametrizations
are related by the transformation $s\to K-s$, where $K$ is the real
period of the elliptic function.} The region $|u|< 2g$ is that of the
``giant magnons'', in the language of \cite{Maldacena-Swanson}, and is
characterized by finite values of the periodic momenta $p$.  The
region $|u|> 2g$ corresponds to the ``plane-wave'' limit \cite{BMN},
where the momenta are of the order $p\sim1/g$.
  
It is more difficult to see what happens exactly at the points $u=\pm
2g$ and in their vicinity.  These points separate the two regimes
discussed above, and they are not properly described by either of the
two parametrizations.  In fact, they correspond to momenta of the
order $p\sim g^{-1/2}$, or to the ``near-flat space'' region, again in
the terminology of \cite{Maldacena-Swanson}.  In some cases almost all
the roots of the Bethe equations are concentrated in this region.  It
is possible to use the elliptic parametrization to obtain the strong
coupling expansion in this regime,\footnote{In this case, the
parametrization is obtained by shifting $s\to \pm K/2+s$.} and it is
clear that the expansion in this region involves powers of $g^{-1/2}$,
or equivalently $1/\lambda^{1/4}$.

In order to obtain the strong coupling limit of the dressing kernel
\cite{BES} we have several options.  One is to take the large coupling
limit term by term in the series defining the dressing phase around
$g=\infty$ \cite{BHL,BES}.  For the giant magnon and plane-wave limits
this works, and this is probably the most straightforward way to
obtained the leading orders.  In the near-flat space limit, however,
as it was noticed in \cite{Maldacena-Swanson}, all the terms of the
series contribute to the leading order in $1/g$.  We would like to
emphasize that, although the results for the leading order of the
anomalous dimensions can be obtained without being particularly
careful about the near-plane wave regime, the corrections will be
crucially determined by it.

A second possibility is to use the representation of the dressing
kernel given by the ``magic formula'' of \cite{BES}, and this
representation was already exploited in \cite{Benna2}.  In this paper
we give another related integral representation
\begin{equation}
K_d(u,u')=4\sum_{n\geq1} \int_{-\infty}^\infty dv\;  
K_-^{1,n}(u,v)\; \(K_+^{n,-1}(v,u')-K_+^{n,1}(v,u')\)\,,
\la{drekint}
\end{equation}
where the kernels $K_\pm(u,u')$ are the odd/even parts of the basic
$su(1|1)$ kernel $K(u,u')$, the inverse Fourier transform counterparts
of the kernels $K_{0,1}(t,t')$ in the BES decomposition \cite{BES}.
The upper indices $K_\pm^{m,n}(u,u')$ mean essentially that the
variables $u$ and $u'$ are defined by elliptic functions of modulus
\begin{equation}
\frac{k'_n}{k_n}=\frac{n}{4g}\;, \qquad \frac{k'_m}{k_m}=\frac{m}{4g}\;.
\end{equation}
Similar variables were used in \cite{BHL} and they are associated to
bound states of magnons \cite{Dorey, Chen:2006gq}.  The sum in
(\ref{drekint}) resembles to a sum over intermediate states; in
particular, the energy associated to these ``intermediate states'' is
of the type \cite{Dorey}
\begin{equation}
E_n =\sqrt{n^2+ 16g^2 \sin^2p/2}\;.
\end{equation}
In the large $g$ limit, the sum over $n$ can be taken with the help of
the Euler-Maclaurin formula and the integrals over the variables $v$
and $z= n/4g$ are simple enough to be done.  After taking the
integrals, we reproduce the first term in the strong coupling
expansion of the dressing kernel \cite{AFS}.  Although technically
this result is not very useful, it can be considered as another
evidence that the weak coupling and strong coupling definitions of the
dressing kernel \cite{BES} agree.

Turning to the explicit results, we obtain here the leading order
result for anomalous dimensions of the highest excited state in the
$su(2)$ and $su(1|1)$ sectors, both for the Bethe ansatz with the
Beisert, Hern\'andez, L\'opez/Beisert, Eden, Staudacher (BHL/BES)
phase, and without it.  Even if the kernel without the dressing phase
is by now only of limited interest, we can still ask the question
whether the $su(1|1)$ sector arises from an effective model, in the
same way the $su(2)$ sector appears from the reduction of the Hubbard
model at half-filling.  The large $g$ limit of the $su(1|1)$ sector
should give a hint for the underlying degrees of freedom.  In the
$su(2)$ sector, which was solved in \cite{RSS,Zarembo05}, the degrees
of freedom are magnons, which become free as $g\to\infty$.  Our result
in the $su(1|1)$ sector shows that these degrees of freedom do not
correspond to a free system, and they are rather difficult to
characterize.  Two thirds of the excitations are of the type ``giant
magnon'', with finite momenta, and one third are of the type ``plane
wave'', with momenta of order $1/g$, see also the numerical results of
\cite{Beccaria-I}.  We find for the energy without the dressing factor
\begin{eqnarray}
E_{su(1|1)}=\frac{8\ln 2}{\pi} g L\;,\qquad E_{su(2)}=\frac{4}{\pi} g
L \, .
\end{eqnarray}
The $su(2)$ energy is the $g\to\infty$ limit of the exact result
\cite{RSS,Zarembo05}.  When taking into account the BHL/BES phase,
almost all roots of the Bethe equations, both for the $su(2)$ and
$su(1|1)$ sector, fall into the region ``near-flat space'' regime,
with momenta of order $1/\sqrt{g}$.  This phenomenon was first
discussed in \cite{AFS}, where the $\lambda^{1/4}$ behavior was
obtained from the BA equations.  The integral equations are able to
reproduce the leading result for the anomalous dimensions of the
corresponding states, which were also obtained numerically and
analytically in \cite{Beccaria-I, Beccaria-II}
\begin{eqnarray}
E^d_{su(1|1)}=\sqrt{2\pi g} \, L\;,\qquad E^d_{su(2)}=\sqrt{\pi g}\,
L\, .
\end{eqnarray}
%
% Although we have not computed the next correction in $1/\sqrt{g}$,
% we are able to see that there are potentially other corrections that
% the ones considered by \cite{Beccaria-II}, in particular because a
% fraction of the solutions may be concentrated in the region ``giant
% magnons''.

We also analyzed the strong coupling limit of the equation derived by
Eden and Staudacher (ES) \cite{ES} for the anomalous dimension of the
twist-two operator without and with the dressing phase (BES equation).
Similarly to the approaches \cite{Benna-I,KL06}, we obtain that the
strong coupling limit of the ES equation is pathological.  The BES
equation has a better behavior, and we are able to reproduce the
leading term of the density obtained recently by Alday {\it et al}
\cite{Benna2} using the Fourier space representation,
\begin{eqnarray}
 \sigma^<(u)=\frac{1}{4\pi g^2}\; 
 \qquad &{\rm for}& \qquad  |u|<2g\, ,
 \\ \no
 \sigma^>(u) =\frac{1}{4\pi g^2}(1-\cosh s/2)\quad 
 &{\rm for}& \quad u=2g\coth s \;, \quad |u|>2g\;.
 \end{eqnarray}
This solution reproduces the leading behavior of the universal scaling
function (\ref{fdeg}).

For the next order in the density, an important fraction of the roots
lie near the points $u\simeq \pm 2g$.  We show that without properly
considering the scattering of the magnons in this region, the density
is non-normalizable.  We leave the fine analysis of the near-flat
space region for future work.

The paper is organized as follows: in section 2 we write down the
integral equations corresponding to the three sectors, in section 3 we
present in detail the elliptic parametrization and we take the strong
coupling limit of one of the the building blocks, the $su(1|1)$
kernel.  In section 4 we compute the energy of the highest states for
the $su(2)$ and $su(1|1)$ without the dressing kernel and discuss the
leading order solution of the Eden-Staudacher equation.  In section 5
we give an integral representation of the dressing kernel and we check
that it reproduces the already known results at strong coupling.  We
also give the strong coupling limit of the AFS phase \cite{AFS}.  The
integral equations with the dressing phase equations are considered in
section 6.

\section{ Bethe ansatz equations in integral form}
\label{bai}
\setcounter{equation}{0}

If we consider highly excited states, with a large number of magnons,
the Bethe ansatz equations can be solved by transforming them into
integral equations.  These integral equations are particularly simple
when the number of magnons is maximal, such that there are no holes in
the magnon distribution\footnote{The integral equation for the highest
state in the $su(1|1)$ sector was written down by one of us and M.
Staudacher, \cite{SS05} and in \cite{ArutyunovTseytlin}, the equation
for the antiferromagnetic state in the $su(2)$ sector was solved in
\cite{RSS}, and the equation in $sl(2)$ sector, or the Eden-Staudacher
equation, was derived in \cite{ES}.}.  Further simplification arise if
we restrict ourselves to one of the three rank-one sectors, $su(1|1)$,
$su(2)$ or $sl(2)$.  The $su(1|1)$ comprises one complex boson $Z$ and
one fermion $U$, and the state with the maximum number of fermions
$$ \Tr U^L$$
is, at least at weak coupling, the one with highest energy.

In the $su(2)$ sector, corresponding to two complex bosons $Z$ and
$X$, the state with highest energy is, again at small coupling $g$,
the anti-ferromagnetic state.  This is, in the case of even length,
the singlet state
$$ \Tr (ZX)^{L/2}+... \;.$$
The $sl(2)$ sector corresponds to combinations of one boson $Z$ and the covariant derivative $D$.  The interesting operators are the so-called twist-$L$ operators with spin $S$,
$$ \Tr D^S Z^L\;.$$

The equations in the three rank-one sectors can be written in the
compact form \cite{BSansaetze, BDS}
\begin{equation}
\label{bag}
\left(\frac{x_k^+}{x_k^-}\right)^L=\prod_{l}^M\frac{1-{g^2/x_k^+x_l^-}}
{1-{g^2/x_l^+x_k^-}}\left(\frac{x_k^+-x_l^-}{x_k^--x_l^+}\right)^\eta\;
e^{i\theta(u_k,u_l)}\;,
\end{equation}
where $\eta=1,0,-1$ for $su(2)$, $su(1|1)$ and $sl(2)$ respectively.
The variables $x^\pm$ are defined by
$$x(u)=\frac{1}{2}u\left(1+\sqrt{1-{4g^2}/{u^2}}\right)\;,   \quad
x^\pm=x(u\pm i/2)\;.
$$
It is often convenient to work with the momentum $p$, the physical
variable of the magnon, defined by
\begin{equation}
\label{utop}
 u(p)=\frac{1}{2}\cot\frac{p}{2} \sqrt{1+16g^2\sin^2{\frac{p}{2}}}\;,
 \qquad {\rm or} \quad x^+/x^-=e^{ip}\;.
\end{equation}
We are going to consider both the case where the phase
$\theta(u_k,u_l)$ is zero and the case where it equals the BHL/BES
dressing phase \cite{BHL,BES}.

Generically, the Bethe equations for the rank-one sectors can be
written as
\begin{equation}
e^{ip_kL}=\prod_{l=1}^M e^{i\varphi(u_k,u_l)}\, ,
\end{equation}
where $\varphi(u_k,u_l)$ is the scattering phase for two magnons.  Let
us exemplify the derivation of the integral equation on the $su(1|1)$
case, which is the simplest one.  We consider for convenience $L$ odd.
The total number of magnons is $M=L$.  When $g=0$, the equations
(\ref{bag}) correspond to a free fermion system, with the occupation
numbers
\begin{equation}
p_k^{(0)}=2\pi k/L\;, \qquad k=-\frac{L-1}{2},\ldots, \frac{L-1}{2}\, .
\end{equation}
When $g$ increases, the momenta start to evolve, according to the
Bethe ansatz equations (\ref{bag})
\begin{equation}
\label{discreteba}
\frac{p_k}{2\pi }=\frac{k}{L} +\frac{1}{2\pi L}\sum_{l=1}^M \varphi(u_k,u_l)\, .
\end{equation}
%
%Due to the fermionic nature of the magnon excitations, the mode
%numbers $n_k$ have to be distinct.  At maximal filling, when the
%number of magnons $M$ equals the length $L$, the only choice is to
%have $n_k=k$.
When the length of the chain is very large, we can take the continuum
limit.  We introduce the variable $t=k/L$ and the density of
rapidities $\rho(u)=-dt/du$.  The derivative with respect to $u$ of
the equation (\ref{discreteba}) gives
\begin{eqnarray}
\label{intgauge}
&\ &\rho(u)=-\frac{1}{2\pi  }\frac{dp}{du}+
\int_{-\infty}^\infty{d u'}\;K(u,u')\  \rho(u')\;,
\end{eqnarray}
where the kernel is the derivative of the scattering phase 
\begin{equation}
K(u,u')= \frac{1}{2\pi  }\frac{d}{du} \;\varphi(u,u')\;.
\end{equation}
The derivation of the $su(2)$ equation is similar, with the difference
that the maximum number of magnons is $L/2$ instead of $L$.  This
derivation follows closely the analysis of the antiferromagnetic state
in the XXX model \cite{faddeev}.  A more sophisticated treatment which
is able to take into account the finite size corrections, was used in
\cite{feverati} to study the antiferromagnetic state in the BDS model
\cite{BDS} and in the Hubbard model.

The Bethe integral equations (\ref{bag}) can be formulated in terms of
three basic kernels.  The first one is the kernel $K(u,u')$,
corresponding to the $su(1|1)$ case without the dressing phase.  The
second one is the $su(2)$ kernel $K_{su(2)}(u,u')$, and the third one
is the dressing kernel, corresponding to the dressing phase
$\theta(u,u')$
\begin{eqnarray}
\label{kernel1}
K(u,u')\ \ &=&\frac{1}{2\pi i}\frac{d}{du}\left( \ln
\left(1-\frac{g^2}{x^+(u)x^-(u')}\right)-\ln\left(
1-\frac{g^2}{x^-(u)x^+(u')}\right)\right)\\
\label{kernel2}
K_{su(2)}(u,u')&=&\frac{1}{2\pi i}\frac{d}{du}\ln \left(
\frac{u-u'+i}{u-u'-i}\right)=-\frac{1}{\pi}\frac{1}{(u-u')^2+1}
\\
\label{kernelc}
K_d(u,u')\ \ &=&\frac{1}{2\pi}\frac{d}{du}\theta(u,u')\;.
\end{eqnarray}
When the dressing phase is taken into account, the complete kernel is
\eqal\totker{\label{totker} {
\cal K}_{\rm tot}(u,u')&=&(1-\eta)K(u,u')+\eta
K_{su(2)}(u,u') +K_d(u,u')\no\\
&\equiv&    \CK(u,u')  +  \eta
K_{su(2)}(u,u')  \;.  }
It will be sometimes convenient to work with the phase associated with
the kernel $\CK$ \eqal\defphi{ \CK(u,u')= \frac{1}{ 2\pi} \p_u \phi
(u, u')\, , \qquad \phi = (1- \eta) \vp + \th.  \la{defphi} }

The $sl(2)$ sector is special in the sense that the number of magnons,
also called spin, $M=S$, is not bounded, even if the length is finite
($L=2$ for the twist-two operator).  The lowest state with $S$ magnons
can be characterized by the numbers
\eqn\nkk{ n_k =k+{\rm sign}(k)\;, \qquad k=-(S-1)/2,\ldots ,(S-1)/2\;.
}
The corresponding integral equation is given, again in the absence of
the dressing phase, by
\begin{equation}
\label{sl2eq}
\rho(u)=\frac{2}{s}\delta(u)+\frac{1}{\pi}\int_{-\infty}^\infty{d u'}\
\frac{\rho(u')}{(u-u')^2+1}+2\int_{-\infty}^\infty{d u'}\;K(u,u')\
\rho(u')\, , 
\end{equation} 
where the term proportional to $\delta(u)$ comes from the distribution
of the mode numbers $n_k$.  The limit $g=0$ of this equation is
singular, and it was solved in \cite{Korchemsky}.  The solution for
large $S$ is
 \begin{equation}
\label{rhozero}
\rho_0(u)=\frac{1}{\pi
S}\ln\frac{1+\sqrt{1-4u^2/S^2}}{1-\sqrt{1-4u^2/S^2}}=\frac{2\ln S}{\pi
S}+\CO(1/S).
\end{equation} 
 Eden and Staudacher \cite{ES} chose to separate the density $\rho(u)$
 into the $g=0$ part, $\rho_0(u)$, and a fluctuation $\sigma(u)$
 \eqn\rhoES{ \label{rhosig}\rho(u)=\rho_0(u)-2g^2\frac{4\ln
 S}{S}\;\sigma(u)+{\cal O}(1/S)=-2g^2\frac{4\ln
 S}{S}\(\sigma(u)-\frac{1}{4\pi g^2}\)+{\cal O}(1/S)\;.  
 }
% The total density is normalized to $1$ $$\int_{-\infty}^\infty
% \rho(u)=\int_{-\infty}^\infty \rho_0(u)=1\;,$$ while the
% normalization of $\sigma(u)$ gives the universal scaling function
% \cite{LipatovPotsdam} \begin{equation}
% f(g)=16g^2\int_{-\infty}^\infty \sigma(u)\;.  \end{equation}
% 
From (\ref{sl2eq}), the equation satisfied by the density fluctuation
$\sigma(u)$ is
\begin{equation}
\label{sl2eqsig}
\sigma(u)=\frac{1}{\pi}\int_{-\infty}^\infty{d u'}\
\frac{\sigma(u')}{(u-u')^2+1}+ 2\int_{-\infty}^\infty{d u'}\;K(u,u')\
\( \sigma(u')-\frac{1}{4\pi g^2}\)\;.
\end{equation} 
This is essentially the equation derived by Eden and Staudacher
\cite{ES}, with the inhomogeneous term written differently.  This way
of writing has the advantage to show that the separation of $\rho(u)$
into $\rho_0(u)$ and $\sigma(u)$ makes sense, at large $g$, if
$K(u,u')$ is sufficiently well behaved at large $u'$.
%\footnote{Let us note that here we work with the true, unsymmetrized
%kernels, whose Fourier transforms are ambiguous at $t'=0$.} such that
%This is certainly the case for the perturbative regime in $g$, but,
%as we will se later, it is not the case for $g\to\infty$ and the
%Eden-Staudacher kernel.
%
As we shall see, if $K(u,u')=\CO(1)$ for large $g$ and $|u'|>2g$, the
equation (\ref{sl2eqsig}) will have only the trivial
solution\footnote{At least if the kernel is non-degenerate.}, since at
leading order in $1/g$ one can write
\begin{equation}
\int_{-\infty}^\infty{d u'}\;K(u,u')\ \( \sigma(u')-\frac{1}{4\pi g^2}\)=0\;.
\end{equation} 
In this case, the separation of the density into a free part and a
perturbation is inconsistent.  The Eden-Staudacher kernel is exactly
of this type, and this might be the reason the strong coupling limit
of the Eden-Staudacher equation is so pathological \cite{KL06,
Benna-I}.  The BES kernel, on the other hand, vanishes at the order
$\CO(1)$ for $|u'|>2g$, and the strong coupling limit of the solution
of the BES equation is well behaved.

\section{ 
Elliptic parametrization of the kernels}

\label{epk}
\setcounter{equation}{0}

\subsection{The elliptic map }

In the strong coupling regime, it is convenient to rescale the
variables $u$ and $x^\pm$ in order to eliminate the overall factors of
$g$.  We will use the rescaled variables throughout the rest of the
paper
\eqn\uhere{
u_{ }=\frac{u_{\rm old} }{ 2g} ,\quad \xpm_{ }
= \frac{\xpm_{\rm old}}{g}, \qquad  \e=\frac{1}{ 4g } \, .
}
The function  
\eqn\defxu{
x(u) = u+ \sqrt{ u^2- 1}
}
has a branch cut along the interval $[-1, 1]$.  Since the Bethe
equations involve the shifted variables $u\pm i\e$, they contain two
symmetric cuts, $[-1 - i\e, 1-i\e]$ and $[-1 + i\e, 1+i\e]$.

 The cuts can be removed by introducing a global
parametrization.  We eliminate the parameter $u$ from
 \eqal\uxpxm{ u_\pm \equiv u \pm {i\e} = \half( x^\pm + 1/ x^\pm) } to
 obtain the following relation between $ \xp$ and $ \xm$ 
\eqn\opom{ \( \xp- \xm\) \(1- \frac{1}{ \xp \xm}\) = 4i\e.
\label{cctr}
}
If we require that the momentum $p$ is real, then $ x ^+$ and $ x ^-$
are complex conjugate and $u$ real.  Then the condition (\ref{cctr})
for $ x = x ^+$ and $\bar x = x ^-$ define a contour $\CC$ in the
complex $ x$-plane.  The contour consists, for $g$ real, of two
connected components, $\CC_\infty$ and $\CC_0$, which are exchanged by
the particle-antiparticle transformation $ x \to 1/ x $.  We denote by
$\CC_\infty $ the component that contains the point $ x =\infty$; the
other component contains the point $ x=0$.  In the limit $\e\to 0$ the
contours $\CC_\infty $ and $\CC_0$ develop cusps where they touch
(Fig.1).  At this point the contour $\CC$ is the union of the real
line and the unit circle.  A second singular value is $\e = \pm i $.
Here the two connected component join into one (Fig.2).

  %%%%%%%%%%%%%%%%%%%%%%%%%%%%%%%%% 
\vskip 30pt
 
\centerline{ \epsfxsize=90pt   \epsfbox{ 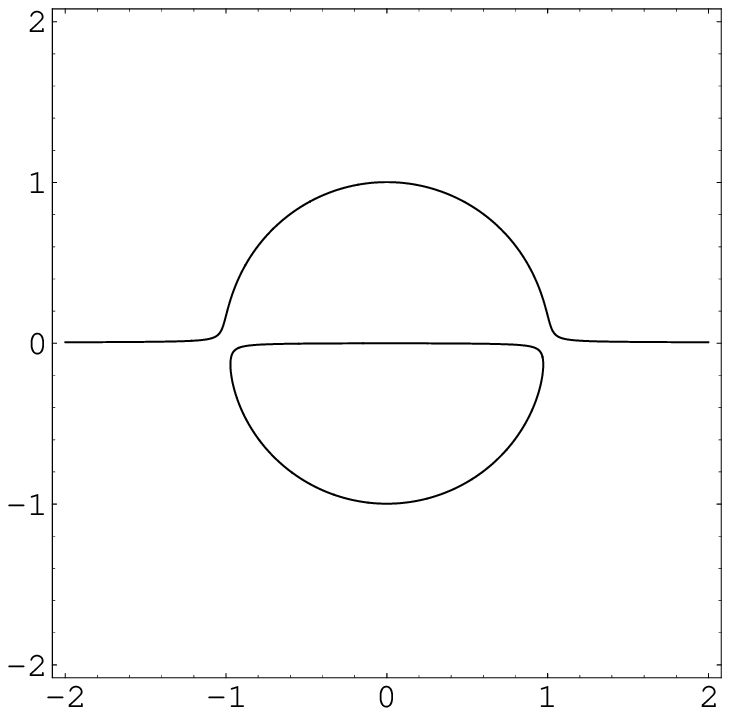   } 
\ \ \  \ \  \   \epsfxsize=90pt \quad  \epsfbox{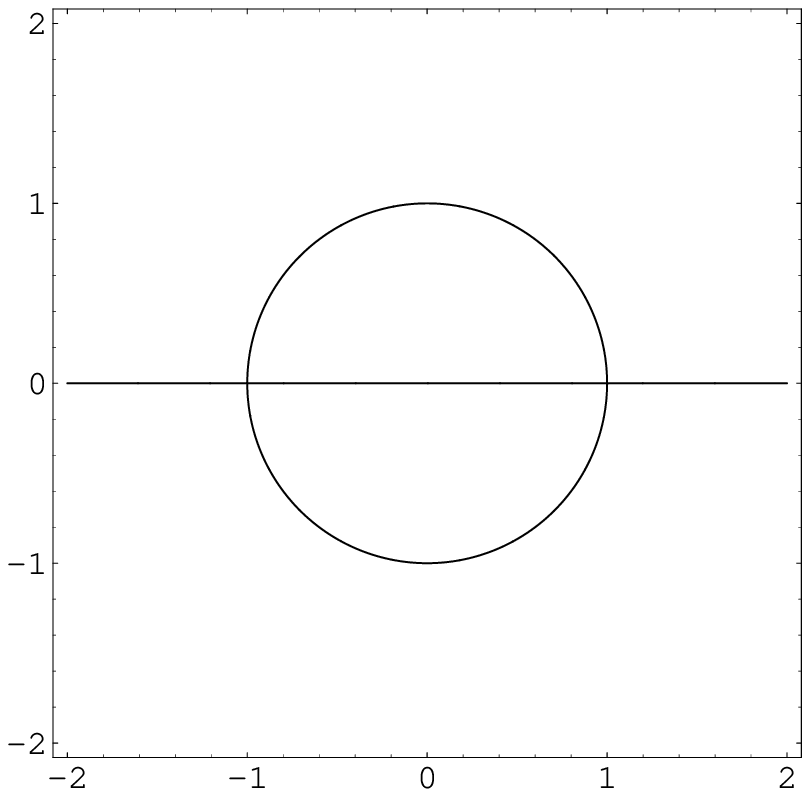   } 
\ \ \ \ \ 
\epsfxsize=90pt 
\quad  \epsfbox{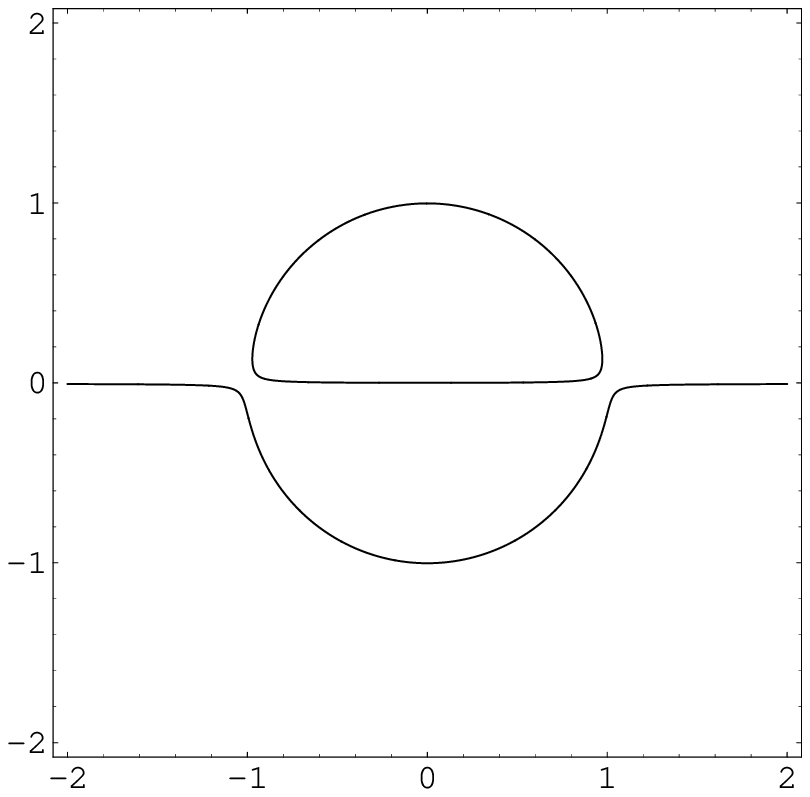   }}
\vskip -3.0cm 
\hskip 2cm $\CC_{\infty}$\hskip 8.9cm $\CC_0$

\vskip 1.4cm \hskip 3cm$\CC_0$\hskip 9cm $\CC_\infty$

\vskip 1.5cm

\vskip -1cm
\centerline{ $\e>0$\qquad \qquad \qquad \quad\qquad$\e= 0$ \quad
\qquad\qquad\qquad\qquad $\e<0$}
%%%%%%%%%%%%%%%%%%%%%%%%%%%%%%%%%%  
\bigskip
\centerline{\small Fig.1 : The contour $\CC$ in the $ x$ plane for
three real values of $\e$ close to 0.}
  
 \vskip  20pt

%%%%%%%%%%%%%%%%%%%%%%%%%%%%%%%%% 
 \epsfxsize=90pt
\vskip 20pt
 \centerline{
\epsfbox{ 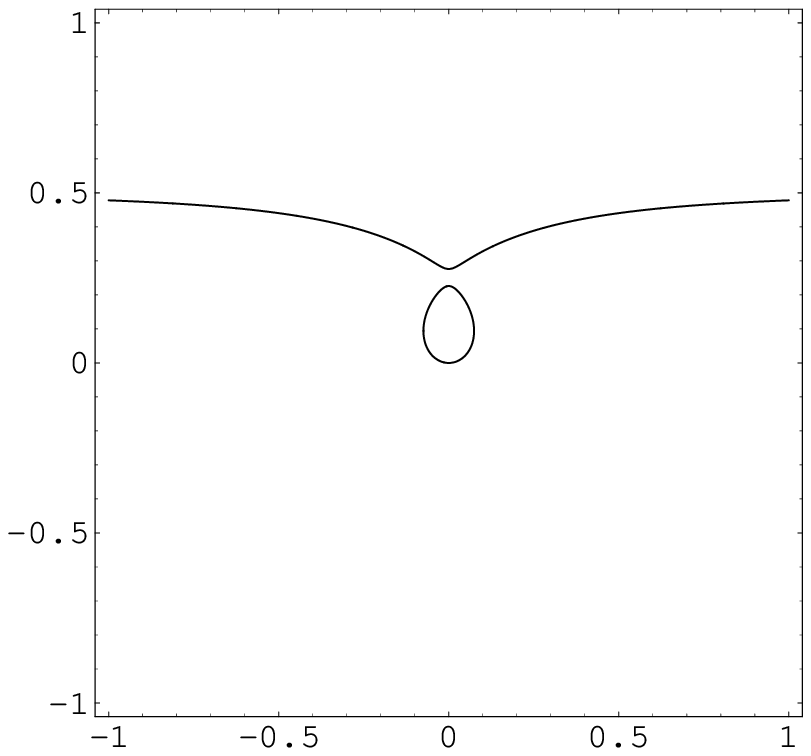   }
\epsfxsize=90pt\quad 
\quad\quad\epsfbox{  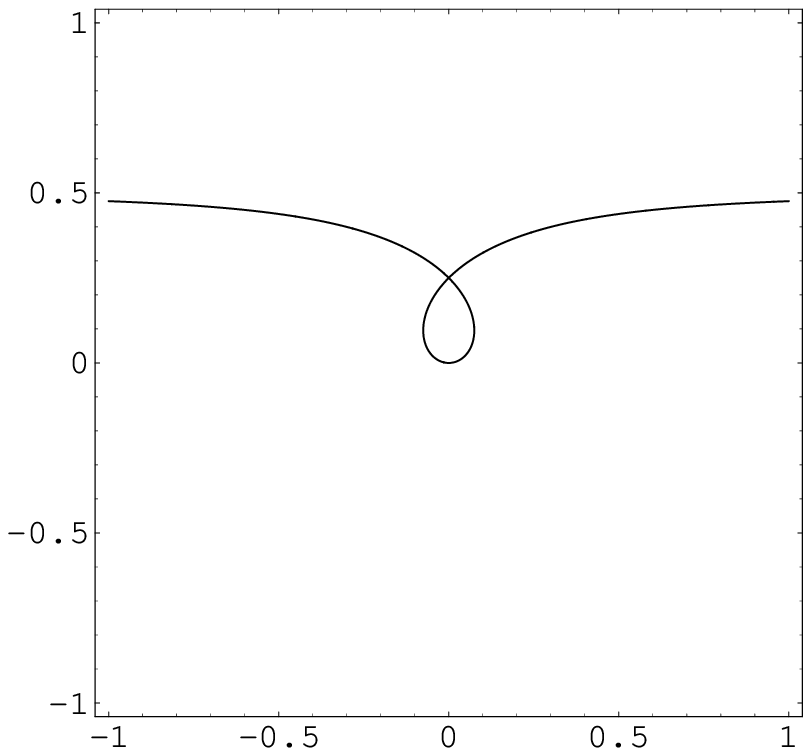   }
\epsfxsize=90pt
\quad\quad \epsfbox{ 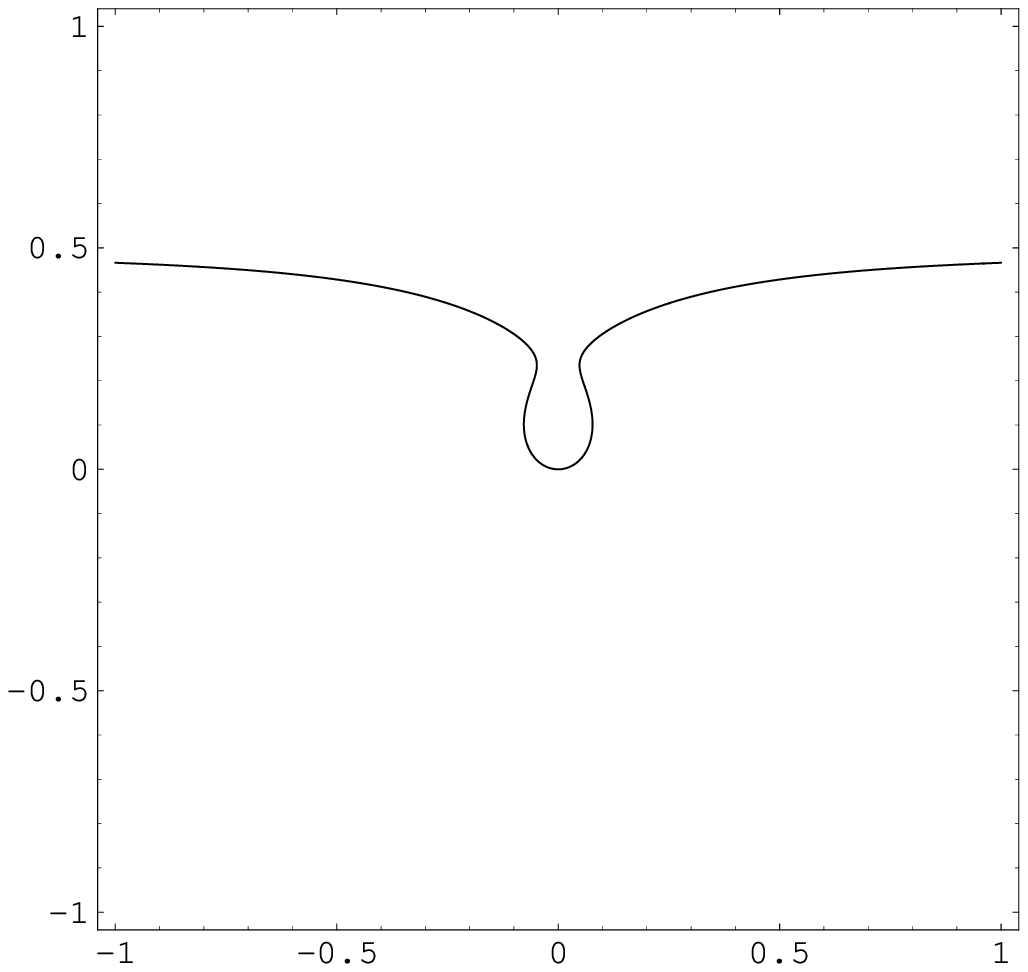   }}
\vskip -5pt
%%%%%%%%%%%%%%%%%%%%%%%%%%%%%%%%%%  
 
\centerline{ \quad $i\e< 1 \quad \qquad$\qquad \quad\qquad$i\e= {1 }
$ \qquad\quad\quad\quad\ \ \ \ \ \ \ $i\e> {1 } $} \bigskip

\centerline{\small Fig.  2 : The contour $\CC$ for three imaginary
values of $\e$ close to $-i$.}
  
 \vskip  20pt
 
\noindent

\noindent

The equation for the contour $\CC$ can be written conveniently in
terms of the momentum $p$ and a complementary parameter $\b$
   \eqn\defpq{ x ^\pm= e^{( \b \pm i p)/2}\, .  
   \la{defpq} 
   } 
In these variables the relation (\ref{cctr}) reads
\eqn\repy{ \sin \frac{p}{ 2} \sinh \frac{\b}{ 2} = {\e} \la{epq}
}
 and the parameter $u$ is expressed as 
 \eqn\upb{ u = \cos\frac{p}{ 2}\cosh\frac{\b}{ 2} \, .  \la{upq}}

In the following, we consider only positive values of $\e$, and
$p\in[0,2\pi]$.  This insures that $u$ in (\ref{upq}) takes all the
real values once and only once.

 The relation (\ref{epq}) is resolved by elliptic map\footnote{Our map
 is related to that of \cite{Janik} by a Gauss-Landen transformation $
 k_{\rm Janik} = 2 \sqrt{k'} /( 1+ k')$.  } with modulus $k$ defined
 as
\eqn\defk{ \e= \frac{k'}{ k} \qquad {\rm or}\qquad k=\frac{1}{\sqrt{1+\e^2}}\, .  }
Then $\b$ and $p$ are parametrized by the Jacobi elliptic amplitude
function
\eqn\pofubis{ p(\s) = \pi - 2\, {\rm am} (\K-\s, k),\ \ \b(\s)= -i\pi
-2i\, {\rm am} (i\K' - \s, k).  }
When the elliptic parameter $s$ sweeps the interval $[0, 2\K]$, the
momentum $p$ increases from $0$ to $2\pi$.  The symmetries of $p$ and
$\b$ are
\eqn\psymm{
p(-s) = - p(s), \quad p(s+2K) = p(s) + 2\pi
}
 \eqn\bsym{ \quad \b(-s ) = \b(s) +2\pi i,\quad \b(2 K-s) = \b(s). 
 }
The functions we will work with are actually expressed in terms of
Jacobi elliptic functions
 \eqal\qofu{ e^ {\pm \b /2} &=& \frac{1\pm \dn \s }{k \, \sn \s} ,
 \quad e^{\pm ip/2} = \cd\s \pm i k' \, \sd \s 
 \cr &&\cr
 \p_\s p &=&
 {2k' \nd \s} , \qquad \ \p_\s \b \ = - 2 \, \cs s .  
 \la{qofu}}
 The contours $\CC_\infty $ and $\CC_0$ are parametrized respectively
 by $s\in I_\infty$ and $s\in I_0$, where
\eqn\Iinfty{
I_\infty = [0,2\K],\quad
 I_0 = [2i\K', 2\K+2i\K'].
}
The original rapidity variable $u$ is parametrized as
\eqn\defU{
    \label{defU}
      \uu (\s) = \frac{1}{ k}\, \frac{\cn\s }{\sn\s \ \dn \s} .}
The map (\ref{defU})  obeys the symmetries 
$ \uu(\s)= -\uu(-\s)= \uu(\s+2\K) =
\uu(\s+2i\K') $, as well as
\eqal\symo{
  \label{symo}
     u(\s) \, u(\K -\s) &=&
% u(\half K -\st) \,   u(\half K +\st)  =
{ 1+\e^2}\, .}
%
%     $$ u(\s) u(s+iK')= 1+\e^2.  $$
The real axis in the $u$-plane is the image of the interval
$I_\infty$.  The functions $ \xpm(\s)$ are periodic in $\s$ with
periods $2\K$ and $4i\K'$ and have the symmetries
\eqal\wusym{
\cr  \xp(s)=  \xp(\K - i \K' -s)
=- \xm(-s),\quad 
    \xpm (\s+ 2i \K')={1\over   \xpm (\s)}
%   \quad
%    \xmp (\s)=   -  \xpm(\s \mp i\K' +\K)
   .}     
The first symmetry preserves the contours $\CC_\infty$ and $\CC_0$,
while the second symmetry exchanges them.  Note that the factors
$e^{\b/2}$ and $e^{\pm i p/2}$ are themselves antiperiodic in
$\s\to\s+2\K$.  This is a little complication, due to the fact that we
are working with trigonometric functions of $p/2$, which are periodic
with period $4\pi$.  However, in the physical variables, they always
appear in combinations which are periodic with period $2\pi$.  As we
already mentioned, we prefer to work with $p$ in the interval
$(0,2\pi)$.  \bigskip
 \vskip 40pt \centerline{ \epsfxsize=190pt \epsfbox{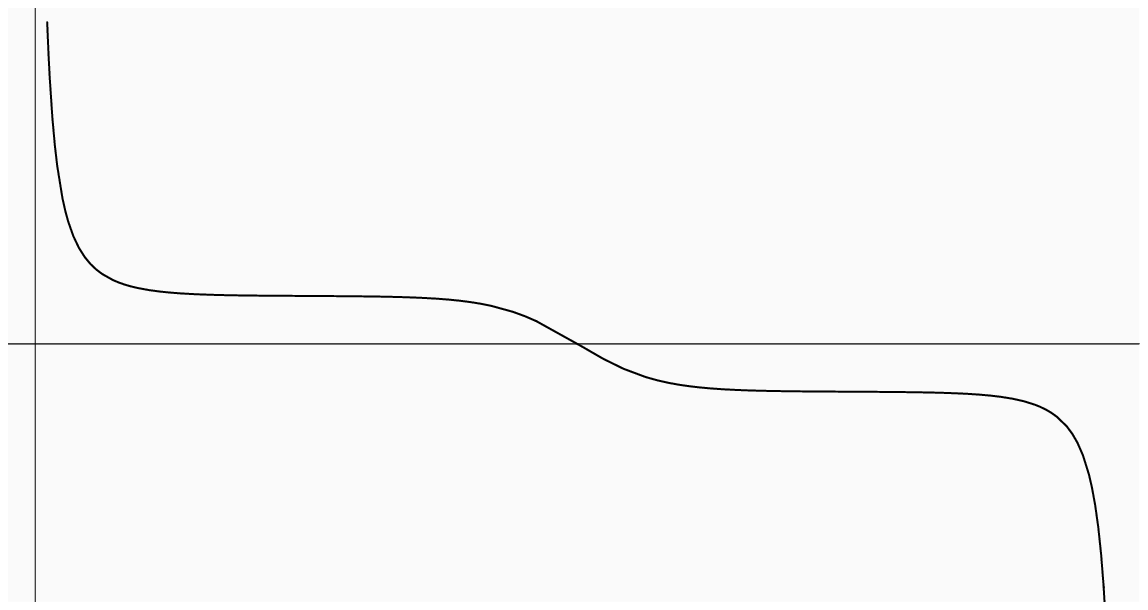} \hskip
 35 pt \epsfxsize=200pt \epsfbox{ 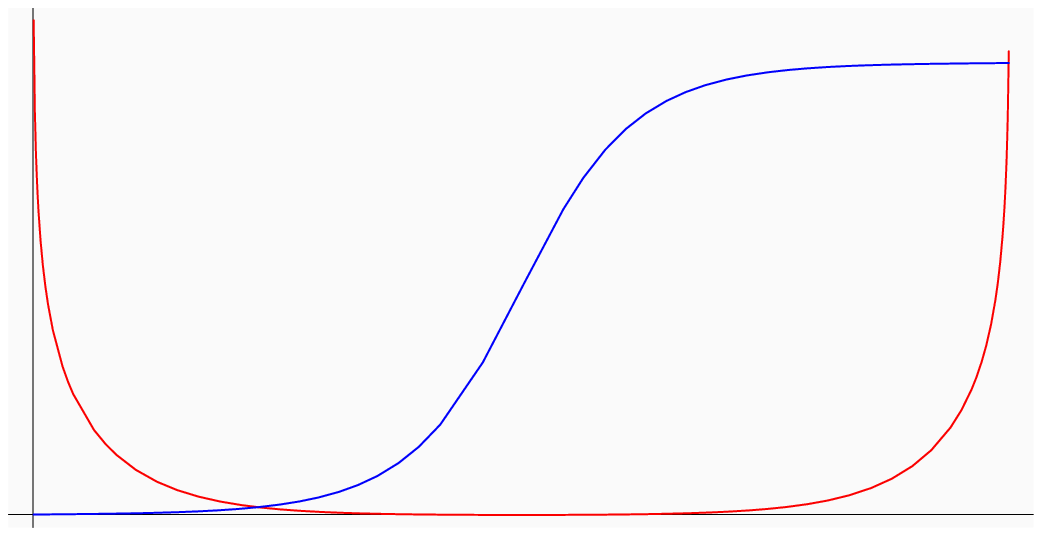 } }
\vskip -3.5cm 
 \centerline{$u(s)$ \hskip 390 pt 
    }
   \centerline{ \hskip 310pt $p(s)$\hskip 60pt $\b(s)$
   }
   \vskip 2.5cm
   \centerline{$0$ \hskip 160 pt $\  2{\rm K}$ \hskip 40pt
 $0$ \hskip 167 pt $ 2{\rm K}$ 
    }
  \begin{center}{\small Fig.  3: Left: The function $u(s)$\ for $\e =
  10^{-3}$ on the interval $[0, 2{\rm K}$];\\
 Right: The functions $p(s)$ (in blue) and $\b(s)$ (in red) for $\e =
 10^{-3}$.  }
 \end{center}
\vskip 20pt

In the limit $\e\to 0$, where $\K\to\infty$ and $\K'\to {\pi/ 2}$, the
elliptic parametrization degenerates\footnote{Another point where the
torus degenerates into a cylinder is $\e=\pm i$, or $g^2=-1/16$, {\it
cf.} Fig.2.} and the variables $x$ and $u$ become hyperbolic functions
of $s$.  As the plot of $u(s)$ in Fig.  3 suggests, the
parametrization interval splits into two different pieces,
corresponding to two different regimes.  By periodicity\footnote{As
already explained, the variables $u$ and $x$ are periodic with period
$2\K$, and we can safely do the shift $s\to s-2\K$ for them, but for
the variables $p,\ \b$ we have to remember that the point $-\hf \K$
was originally ${3\over 2} \K$.}, we can choose the parametrization
interval as $[-\hf \K, {3\over 2} \K]$; then the first regime
corresponds to the interval $[-\hf \K, \hf \K]$, which becomes the
whole real axis when $\K\to \infty$.  The second regime, corresponding
to the interval $[\hf \K, {3\over 2} \K]$ can be mapped also to the
interval $[-\hf \K, \hf \K]$ by the transformation
\eqn\sstilde{ s\to \K-s.
 \label{sstilde}
 }
The first interval corresponds to values $|u|>1$, while the second
corresponds to $|u|<1$.  We therefore introduce the suggestive
notation
   \eqn\defIm{ I^{_>}=[-\half \K,\half \K] \, ,\qquad \quad I
   ^{_<}= [\half \K, \textstyle{3\over 2}\K] }
and use the asymptotic formulas valid in the limit $\e\to 0$ and for
$s\in I^{_>}$ 
 \eqal\trigSSN{ \sn \s &= \tanh \s
 % +  {k'^2 (\sinh \s \cosh \s - \s) \over 4 \cosh ^2\s } 
,\qquad \qquad \qquad 
   \ \ \  \sn(\K- \s )&=
1  \nonumber\\
   \cn \s  &=
 {1 /\cosh \s }
 %  \(1-{ k'^2\over 4}   \tanh \s (\sinh \s \cosh \s - \s)  \)
 ,\qquad \qquad \qquad  \cn(\K-\s ) &= k' \sinh \s ,
  \cr
  \dn \s    &=
   {1/ \cosh \s } 
  %\(1+{ k'^2\over 4}   \tanh \s (\sinh \s \cosh \s +\s)  \)
  ,\qquad \qquad \qquad 
   \dn(\K-\s  ) &= k' \cosh \s 
  %\cd (K+\s') &=& - \sn \s'\approx  - \tanh \s'
\label{trigSN}
. } 
The next terms in the expansion in $\e$ are given in Appendix A.

In the hyperbolic limit, the expression (\ref{defU}) for $u$ becomes,
in the two different regimes,
\begin{equation}
\label{ugup}
u^{_>}(s)= u(\s) \simeq \coth s\;, \qquad u^{_<}(s) = u(\K-\s )
\simeq\tanh s\; .
\end{equation}

The points $s=\pm \K/2$, which are the fixed points of the
transformation $s\to \K-s$, correspond to the points
$u=\pm\sqrt{1+\e^2}$.  In the strong coupling limit, $\e \to 0$, these
are the points $u=\pm1$, and they correspond to meeting points $s
\to\pm\infty$ of the two parametrizations (\ref{ugup}).  These points
will be very important, and we need a better parametrization for them.
In the strong coupling limit the vicinity of these points is
parametrized by (Appendix A)
  \eqn\uKtwo{ u ( \pm \half \K +s)= \pm 1 - \e \sinh 2s + \CO(\e^2)
  .  \la{K/2u}}
The expressions for the momentum $p$ and its counterpart $\b$ in these
regimes are
\eqal\Ktwob{ \sinh \b/2=
%=
% \sinh   {\b(\pm \half \K + \y) \over 2}
   \sqrt{\e}\, e^{\mp s} \qquad 
\sin  p /2=
%= \sin  {p(\pm \half \K + \y) \over 2}
  \sqrt{\e}\,  e^{ \pm s} 
\;,
\la{K/2pb}
 }
 where the upper sign correspond to momenta close to $0$ and the lower
 sign to momenta close to $2\pi$.  A similar parametrization was used
 in \cite{Maldacena-Swanson}.  
% Note that near these points the
% corrections are of order $\e$ and not $\e^2$.

% The Bethe equations are invariant under $\xpm \to -\xpm$.  This
% symmetry exchanges the contours $ \CC_\infty$ and $ \CC_0$ and their
% mirror images with respect ti the real axis, $\bar \CC_\infty$ and
% $\bar \CC_0$.
%    % Note that the contours   $\CC_\infty$ and  $\CC_0$ are    mapped to the 
%% same line $\Re u = i\e$ in the $u$-plane.
%
% Let us finally mention that the Bethe equations can be formulated as
% the conditions for equilibrium of a system of charges in the
% $x$-plane, confined to live on the contour $\CC_\infty$, and their
% images with respect to symmetries $x\to -x$ and $x\to .  Take the
% phase factor \eqn\defOm{ \O(\s,\s') = e^{i\vp(\s,\s')}= \frac{1-
% {1\over x_+(\s) x_-(\s') } }{ 1- {1\over x_-(\s) x_+(\s')}}
% \la{defOm} .} Its derivative is given by \eqn\defK{ {1\over 2\pi}
% \p_\s\vp(\s,\s')={1\over 2\pi i} \p_s\log \O(\s,\s') =- {p'(\s)\over
% 4 \pi } + \tK(\s,\s'), \la{defK}}
 %

\subsection{Strong coupling limit of the  $su(1|1)$  kernel}

We saw that in the strong coupling limit the integration interval
splits naturally into two domains, $|u|<1$ and $|u|>1$, which are the
images of the intervals $I^{_<}$ and $I^{_>}$ in the $s$
parametrization.  As it is clear from Fig.  1, in the first domain the
complex variable $x=\xp$ becomes asymptotically unimodular, while in
the second domain it becomes asymptotically real.  For $\e\to 0$ the
relations (\ref{epq}) and (\ref{upq}) give
\begin{eqnarray}
\beta\to 0\;, \quad u&=&\cos p/2 \qquad\ \ {\rm if} \quad |u|<1\;,\\
\ \ p\to 0 \;, \quad u&=&\cosh \beta/2\qquad {\rm if} \quad  |u|>1\; .
\end{eqnarray}
%
%These equations have been instrumental to obtain the anomalous
%dimensions perturbatively \cite{ES} \cite{AT}.  The purpose of the
%present work is to analyse these equations at strong coupling and to
%test the predictions made by \cite{GKP}.
%
%
Passing to the elliptic parameter $s$ we write the integral equation
(\ref{intgauge}) as
\eqn\Inteqes{ \ \rho(s)=\frac{1}{2\pi }p'(s) + \int_{ I_\infty}{d
s_1}\;K(s,s_1)\ \rho(s_1)\;,
\label{Inteqes}
 }
 where 
\eqn\defthoKs{ \rho(s) = |u'(s)| \, \rho(u), \qquad K(s, s_1) =
 |u'(s)| \, K(u, u_1).  \la{defrhoKs} }
The first term in the r.h.s. of  (\ref{Inteqes}) changes sign because the derivative $u'(s)$ is
negative.  The kernel is equal to the derivative
 \begin{eqnarray}
 K(s,s_1)&=&-\frac{1}{2\pi i}\frac{d}{ds}
 \left[\ln \left(1-e^{-\hf (\beta+\beta_1)}e^{-i{\hf (p-p_1)}}\right)
 -\ln \left(1-e^{-\hf (\beta+\beta_1)} e^{i \hf {(p-p_1)}}\right)\right]
  \cr
  &&\cr
 &=&\frac{1}{4\pi} p'(s)- \frac{1}{4\pi}
 \frac{p'(s)\sinh\hf (\beta+\b_1)
  -\beta'(s) \sin\hf (p-p_1)}
 {\cosh\hf (\beta+\b_1)-\cos\hf (p-p_1)}.
 \label{Kssone}
  \end{eqnarray}
Let us mention here that the piece $p'(s)/4\pi$ has a physical
meaning: it corresponds to a change of the effective length of the
chain.  Every magnon increases the effective length of the chain by
$1/2$.  This is particularly clear on the discrete equations, where,
at strong coupling, the Bethe equations become, as it was noticed in
\cite{ArutyunovTseytlin}
\begin{equation}
e^{ip_k(L+M/2)}=1\, ,
\end{equation} 
under the condition that all $p_k$ are finite.

It is now easy to take the strong coupling limit using the asymptotic
expressions for $p(s)$ and $\b(s)$, which can be found in Appendix A.
The asymptotic form of the kernel is given by four different analytic
expressions depending on whether its arguments are in the interval $
I^{_>}$ or $ I ^{_<}$.  After applying the redefinition
(\ref{sstilde}) to the arguments that belong to the second interval,
we can write the result as a $2\times 2$ matrix kernel
\eqn\pmatrix{
  \begin{pmatrix}
  \Kpp(\s,\s_1)   &  \Kpm(\s,\s_1)   \\
 \Kmp(\s,\s_1)      &  \Kmm(\s,\s_1) 
\end{pmatrix} 
\ :=\ 
 \begin{pmatrix}
  K (\s,\s_1)   \ \ \ \ \  &   K(\s,\K-\s_1) \\
 K  (\K-\s,\s_1)     &\ \  \ \  \ \ K(\K-\s,\K-\s_1) 
\end{pmatrix}
\, .
\la{ppmatrix}
}
The arguments $s,s_1$ are defined in the interval $ I ^{_>}$, which
extends to $[-\infty,\infty]$ when $\e\to 0$.  Similarly the density
splits into two components,
\eqn\defden{
 \begin{pmatrix} 
 \rhp(s)\\
 \rhm(s)
 \end{pmatrix} =
 \begin{pmatrix} 
 \rho(s)\ \ \ \  \ \\
 \rho(\K-s)
 \end{pmatrix}
  }
satisfying, in the limit $\e\to 0$, the normalization
condition\footnote{This is true only for the kernel without the
dressing phase, at the leading order.}
 \eqn\normpp{
  \int_{-\infty}^\infty d s \rhm  +  \int_{-\infty}^\infty d s \rhp =1.
  \la{normpp}
  }
Eq.  (\ref{Inteqes}) now takes the form 
\eqal\matrixeq{\rhp (s) &=&\ \
\int_{-\infty}^\infty ds_1\Kpp(s,s_1)\, \rhp(s_1)  +
\int_{-\infty}^\infty ds_1\Kpm(s,s_1)\, \rhm(s_1) \, ,\\
\no
\\
\rhm (s) &=&
{ p'(s) \over 2\pi}+ \int_{-\infty}^\infty ds_1\Kmp(s,s_1)\,
\rhp(s_1)  + \int_{-\infty}^\infty ds_1 \Kmm(s,s_1)\, \rhm(s_1).
\label{IntM}
  }
The matrix elements of the kernel (\ref{ppmatrix}), evaluated in
Appendix A, are
 \begin{eqnarray}
  \Kmm(s,s')&=&\frac{1}{2\pi}\, \frac{1}{\cosh s}-\delta (s-s')\;, \no \\
  \no \\  
  \Kmp(s,s')&=& \frac{1}{2\pi }\, \frac{1}{\cosh s}- \frac{1}{2\pi }\,
  \frac{1}{\cosh (s-s')}\;, \no \\
  \Kpm(s,s')&=&\frac{1}{2\pi }\, \frac{1}{\cosh (s-s')}\;, 
  \no \\
  \Kpp(s,s'')&=&0\;.
  \label{Kss}
 \end{eqnarray}
Note that in the hyperbolic limit all four integration kernels depend
on the difference of the arguments.  This is a considerable
simplification, since the equation can be now solved using Fourier
transform.
 
In the third regime, which corresponds to the points around $u=\pm 1$
the momenta are parametrized by (\ref{K/2pb}).  This regime will be
very important for evaluation of the density with the BHL/BES kernel.
The appropriate parametrization is given by (\ref{K/2u}) and
(\ref{K/2pb}).
%  \be gin{equation} u=\pm \(1-\half \e \sinh 2s\), \end{equation} and
%  \begin{eqnarray} \sin \half p = {\sqrt{\e}}\;\exp(\mp s) ,\qquad
%  \sinh \half \beta= {\sqrt{\e}}\;\exp(\pm s)\;.  \end{eqnarray}
The only component of the undressed kernel that survives at the leading order is
when either $u, u'\simeq 1$ or $u,u'\simeq -1$
 \begin{eqnarray}
&& K^{++}(s,s') = K^{--}(-s,-s') \\ \no && =- {1\over 4 \pi}\,
\frac{\sinh \left(s-s'\right)+1}{ \cosh \left(s-s'\right) \cosh
\left(s+s'\right)-\sinh \left(s+s'\right) }+\CO(\sqrt{\e})\, .
%  -\frac{e^y \sqrt{\epsilon }}{4 \pi }
\la{KhfK}
\end{eqnarray}
%$$ K^{++}(y,y') = \frac{\sinh \left(y-y'\right)+1}{\pi \left(\cosh (2
%y)+\cosh \left(2 y'\right)-2 \sinh \left(y+y'\right)\right)} $$ $$
%K^{--}(y,y') = \frac{1-\sinh \left(y-y'\right)}{\pi \left(\cosh (2
%y)+\cosh \left(2 y'\right)+2 \sinh \left(y+y'\right)\right)} $$
If only one of the arguments is in the intermediate region, the kernel
is of order $\sqrt{\e}$.
  
One might argue that there are many intermediate regimes, where $p$
scales like an arbitrary
power of  $\e$, 
\begin{equation}
p\sim \e^{1-\gamma}\;, \qquad \beta \sim \e^\gamma\, \quad \gamma \in (0,1)\;.
\end{equation}
However, if $\gamma\neq 1/2$, one of the variables $p$ and $\b$
dominates the other, and these regimes are properly taken into account
in the regions $_<^>$.  Only for $\gamma=1/2$ the quantities $p$ and
$\b$ are of the same order of magnitude and this is why we have to
consider this case separately.

\section{Solving the integral equations in the strong coupling limit
without the dressing kernel}
\label{sclsol}

\setcounter{equation}{0}

\subsection{The $su(2)$ case}

The $su(2)$ kernel is the only one which is of difference form,
therefore it can be exactly solved for any value of $g$ \cite{RSS,
Zarembo05}.  The energy of the antiferromagnetic state is identical to
the energy of the ground state for the Hubbard model at half filling
\cite{LW}.  Here, we would like to comment on the strong coupling
limit of this solution.  The integral equation is particularly simple,
since the kernel becomes simply the delta function
\begin{equation}
\label{su2strong}
K_{su(2)}(u,u')=-\lim_{\e\to
0}\frac{1}{\pi}\frac{2\e}{(u-u')^2+4\e^2}=-\delta(u-u')\, .
\end{equation}
The integral equation reduces then to
\begin{equation}
2\rho(u)=-\frac{1}{2\pi}\frac{dp}{du}\;,\qquad{\rm or} \qquad
\rho(p)=\frac{1}{4\pi}\;.
\end{equation}
The solution corresponds to $L/2$ roots distributed uniformly between
$p=0$ and $2\pi$.  In this case, the density is normalized to $1/2$,
because the total number of magnons is $L/2$.  All roots of the Bethe
equations are inside the interval $|u|<1$ and the energy is
\begin{equation}
E_{su(2)}=4gL\int_0^{2\pi} dp\;|\sin p/2| \; \rho(p)=\frac{4gL}{\pi}\;.
\end{equation}

Physically, the excitations are the (usual) magnons.  The solution
obtained above shows that, in the strong coupling limit, the magnons
are {\it free}, except for the statistical repulsion
(\ref{su2strong}).  The situation is very much similar to that of the
Haldane-Shastry model \cite{Haldane,Shastry}, where the magnons are
also free up to the statistical repulsion, translated into the rule
that two magnons cannot occupy successive momenta.  Let us remind that
the scattering phase for particles with purely statistical interaction
is \cite{Sutherland}
\begin{equation}
\label{statint}
\varphi(p,p')=(\lambda-1)\,\pi\; {\rm sign} (p-p')\;,
\end{equation}
which is consistent with (\ref{su2strong}), the statistical parameter
of the magnons being $\lambda=2$.  The only difference with the
Haldane-Shastry spin chain is that here the magnons have the
dispersion relation
\begin{equation}
E(p)=4g |\sin p/2|\, ,
\end{equation}
which is typical for a finite-difference hamiltonian, while the
Haldane-Shastry magnons have the dispersion relation
\begin{equation}
E_{HS}(p)= p(2\pi-p)\;.
\end{equation}
A natural candidate for a model with purely statistical interaction
and with trigonometric dispersion relation is the Ruijenaars-Schneider
model \cite{RuijSch}.  This suggests the existence of a spin model
which would describe the spin sector of the half-filled Hubbard model
for any value of $g$ and which might be a multi-spin deformation of
the Inozemtsev model \cite{Ino,SS04}.

\subsection{The $su(1|1)$ case}

The solution of the $su(1|1)$ sector is considerably more involved,
but its leading order still can be obtained in closed form.  Since
now the roots of the Bethe equations occur both in the regions $|u|<1$
and $|u|>1$, the integral equation splits up into two coupled
equations for the densities in the two regions 
   \begin{eqnarray}
  2\rhm(s)&=&\frac{3}{2\pi}\frac{1}{\cosh
  s}-\frac{1}{2\pi}\int_{-\infty}^\infty ds'\, \frac{\rhp(s') }{\cosh
  (s-s')}\, ,\\
 \rhp(s)&=&\frac{1}{2\pi }\int_{-\infty}^\infty ds'\, 
  \frac{\rhm(s') }{\cosh (s-s')}\;.
 \end{eqnarray}
 Substituting the second equation into the first we obtain a new
 equation where the kernel is of difference form
\begin{equation}
\rhm(s)=\frac{3}{4\pi}\frac{1}{\cosh
s}-\frac{1}{4\pi^2}\int_{-\infty}^\infty ds'\;
\frac{s-s'}{\sinh(s-s')}\; \rhm(s')\;,
\end{equation}
and therefore is solved by Fourier transformation.  In Fourier space
the equation becomes algebraic,\footnote{In order to keep the notation
simple, we are going to use the same symbol for the Fourier transforms
$\rhm(t)$, $\rhp(t)$ of the densities.  }
\eqn\gpa{
 \rhm(t)={3\over 4\pi}{\pi\over \cosh (\pi t/2)}
-{1\over 4\pi^2}{\pi^2\over 2\cosh^2 (\pi t/2)}   \rhm(t)\;,
 }
  with the solution 
\eqal\Fourrhm{ \rhm(t)&=&\frac{ 6 \cosh (\pi
  t/2)}{1+8 \cosh^2 (\pi t/2)}.  }
Transforming back to the $\s$-variable, we get
\begin{eqnarray}
\rhm(s)&=&\frac{1}{\sqrt{2}\pi}\frac{\cos 2as/\pi}{\cosh s}\;,\\
 a\ \ &=& 
\ln\sqrt{2}\  \simeq .34657\;.
\end{eqnarray}
For the density of roots $\rho(u)$ we obtain in the interval $|u|<1$
\begin{equation}
\label{rhm}
\rhm(u)= {\rhm(s)\over |u'(\s)|}=\frac{1}{\sqrt{2}\pi}\,
\frac{1}{\sqrt{1-u^2}}\cos\left( \frac{a}{\pi} \ln\frac{1+u}{1-u}\right)\;.
\end{equation}

To evaluate  the function $\rhp(u)$, we make the change of variable
$u=\coth s$, which is appropriate for $|u|>1$, so that we get
\begin{equation}
\rhp(s) =\frac{ 1}{2\sqrt{2}\pi^2} \int_{-\infty}^\infty
\frac{ds'}{\cosh s'}\;\frac{ \cos (2as'/\pi)}{\cosh(s-s')}
=\frac{1}{\pi}{\sin (2as/\pi)
\over \sinh s}
\;,
\end{equation}
or, in the initial variables,
\begin{equation}
\label{rhp}
\rhp(u) =\frac{1}{\pi}\frac{1}{\sqrt{u^2-1}}\sin\left( \frac{a}{\pi}
\ln\frac{u+1}{u-1}\right)\;.
\end{equation}
The result shows a singularity at the points $u=\pm 1$; moreover, it
can take negative values.  This happens at values of $1-u\simeq
10^{-5}$ and is presumably due to the roots around $u=\pm 1$ which
were not taken properly into account.  Except for these oscillations,
the result reproduces nicely the solution obtained by numerical
integration of the Fredholm equation\footnote{We thank M. Staudacher
for supplying us the solution obtained by numerical integration.}.

The total normalization of the density is insured, since $\int_{<}
du\; \rho_<(u)=2/3$ and $\int_{>}du\; \rho_>(u)=1/3$.  So two third of
the roots of the Bethe equations fall in the region $|u|<1$ and one
third in the intervals $|u|>1$.  This phenomenon is also present in
the numerical solution of \cite{Beccaria-I}.  Finally, the energy of
the state is given by
\begin{eqnarray}
E_{su(1|1)}&=&4gL\int_{-1}^1du\; \rho_<(u)\; \sqrt{1-u^2}\\
&=&\frac{2\sqrt{2}gL}{\pi} \int_{-\infty}^\infty ds\; 
\frac{\cos 2as/\pi}{\cosh^2 s}=\frac{16a}{\pi}gL\;.
\end{eqnarray}
This result is consistent with that of Beccaria and Del Debbio
\cite{Beccaria-I}.  We obtain for their constant $c_L$ the value
$c_L=2\ln(2)/\pi^2\simeq .14046$. %
\begin{center}
\includegraphics[scale=0.6]{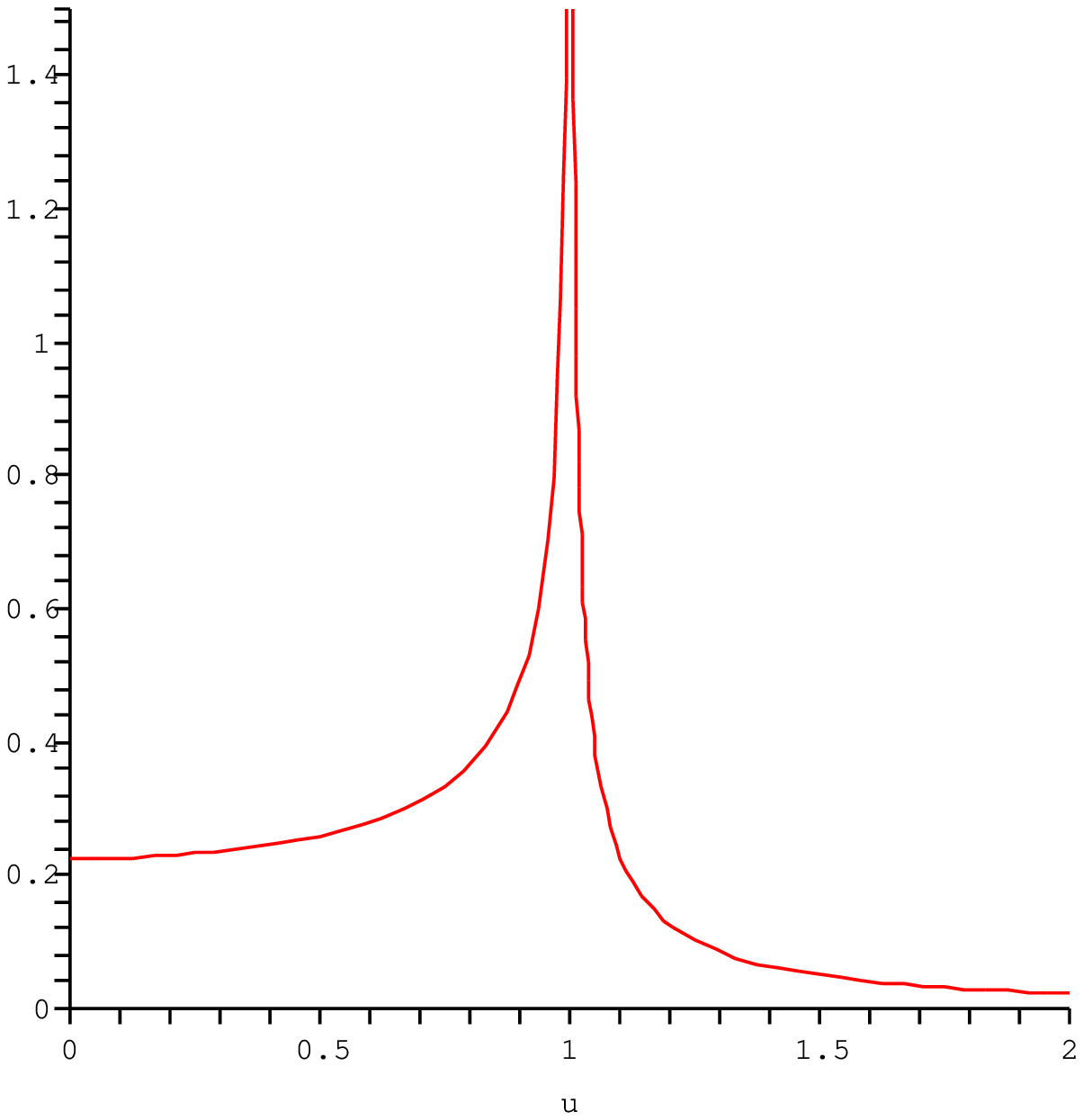}
\end{center}
\begin{center}
{\small Fig.  4: Density profile predicted by the analytic solution
(\ref{rhm}), (\ref{rhp}).  Only the $u>0$ part is shown.  The
oscillations do not show yet at the chosen scale.  }
\end{center}

\subsection{The $sl(2)$ case}

The dimension of the twist two operator was originally supposed to be
determined by the solution of Eden-Staudacher equation \cite{ES}
\begin{equation}
\label{eqnsl2shift}
\sigma(u)=
\frac{1}{\pi}\int_{-\infty}^\infty du'\;\frac{2\e}{(u-u')^2+4\e^2}\; 
\sigma(u')+2\int du'\; K(u,u')\; \left(\sigma(u')-
\frac{1}{2\pi  g} \right)\; .
\end{equation}
%
% %
%\eqn\eqnsltwo{ \label{eqnsl2} \sigma(u)= \frac{1}{2\pi
%g}\frac{d}{du}\left[\frac{1}{x^+(u)}+\frac{1}{x^-(u)}\right]
%+\frac{1}{\pi}\int_{-\infty}^\infty du'\;
%\frac{2\e\;\sigma(u')}{(u-u')^2+4\e^2}\\+2\int du'\; K(u,u')\;
%\sigma(u')\;.  }
%
The anomalous dimension is simply given by the normalization of the
density fluctuation $\sigma(u)$
\cite{LipatovPotsdam}
\begin{equation}
f(g)=16g^2\int_{-\infty}^\infty du\; \sigma(u) \;,
\end{equation}
or alternatively by the formula \cite{ES}
\begin{equation}
f(g)=8g^2\left(1-2g\int_{-\infty}^\infty du\;
\sigma(u)\left(\frac{i}{x^+(u)}-\frac{i}{x^-(u)}\right)\right)\; .
\end{equation}
%
%As explained in section (\ref{bai}), an alternative presentation of
%the Eden-Staudacher equation is
%%
%\begin{equation} \label{eqnsl2shift} \sigma(u)=
%\frac{1}{\pi}\int_{-\infty}^\infty du'\;\frac{2\e}{(u-u')^2+4\e^2}\;
%\sigma(v)+2\int du'\; K(u,u')\; \left(\sigma(u')- \frac{1}{2\pi g}
%\right)\;, \end{equation}
%
As explained in section \ref{bai}, in the limit $\e=1/4g \to0$
the equation (\ref{eqnsl2shift}) becomes
\begin{equation}
\label{sl2simpl}
\int du'\; K(u,u')\; \left(\sigma(u')-\frac{1}{2\pi g} \right) =0\;.
\end{equation}
%
%The manipulation nedded to pass from the equation (\ref{eqnsl2}) to
%involve an shift by infinite constant, since $$\int_{|u'|>1} du' \Kmp
%(u,u')=\frac{1}{2\pi}\int_{-\infty}^\infty \frac{ds'}{\sinh^2
%s'}\(\frac{1}{\cosh s}-\frac{1}{\cosh (s-s')}\)\;=\infty$$
Obviously, this equation has as solution\footnote{Here we use the
variable $u$ rescaled by $2g$, whence the extra factors of $2g$ in the
density.}
\begin{equation}
\label{sigES}
\sigma(u)=\frac{1}{2\pi g}\;
\end{equation}
on the whole real axis.  This solution, first obtained in \cite{ES},
seems to be correct in the interval $|u|<1$, since it insures the
vanishing of the $\CO(g^2)$ term in $f(g)$
\begin{equation}
f(g)= 8g^2\(1-4g\int_{-1}^1
du\;\sigma^{_<}(u)\sqrt{1-u^2}+\CO(1/g)\)=\CO(g)\;.
\end{equation}
In the interval $|u|>1$, although it has the right scaling in $g$, the
solution (\ref{sigES}) is not integrable and it gives formally an
infinite value for the anomalous dimension of the twist-two operator.
As mentioned in the section \ref{bai}, the density (\ref{sigES})
exactly compensate the leading order in $S$ in $\rho_0(u)$ {\it
everywhere} in $u$
\begin{equation}
\rho_0(u)=\frac{4g}{\pi}\frac{\ln S}{ S}+\ldots\;.
\end{equation}
Unfortunately, as we show below, (\ref{sigES}) is the unique solution
to the ES equation at strong coupling. We can write the equations as
\eqal\eqspm{ \frac{1}{2\pi}\int d\s '\; \frac{\sr^<(\s
')}{\cosh(s-s')}\; &=&\frac{1}{8\pi g} {1\over \cosh^2(\s/2)}\, , \\ \no
\frac{1}{2\pi} \int d\s '\; \frac{ \sr^>(\s ')}{\cosh(s-s')}\; +
\sr^<(\s ) &=& \frac{1}{4\pi g} {1\over \cosh^2(\s)} + \frac{1}{2\pi}
{A\over \cosh(\s)} \;, }
where $A=\int ds\;(\sr^<(\s )+ \sr^>(\s ))$ is the normalization of
the density.  In Fourier transformed form, the equations become
 \eqal\eqspmf{ \frac{\sr^<(t)}{2\cosh \pi t/2}\; &=&\frac{t}{2g\sinh
 \pi t} \, ,\\ \no \frac{\sr^>(t)}{2\cosh \pi t/2}\; + \sr^<(t ) &=
 &\frac{t}{4g\sinh \pi t/2} + \frac{A}{2\cosh \pi t/2} \;.  }
The first line is readily solved as
\begin{equation}
\sr^<(t)=\frac{1}{2g\sinh \pi t/2}\, ,\quad {\rm or}\quad
\sr^<(s)=\frac{1}{2\pi g}\frac{1}{\cosh^2s}\, ,\quad {\rm or}\quad
\sr^<(u)=\frac{1}{2\pi g}\;,
\end{equation}
while the solution for the second is 
\begin{eqnarray}
\sr^>(t)=A-\frac{t}{2g\pi \tanh \pi t/2}\, ,\quad &{\rm or}&\quad
\sr^>(s)=A\;\delta(s)+\frac{1}{2\pi g}\frac{1}{\sinh^2s}\, ,\no \\
{\rm or}\qquad   \sr^>(u)&=&-A\;\delta(u-\infty)+\frac{1}{2\pi g}\;.
\end{eqnarray}
The normalization $A$ is not fixed by the equation.

All other approaches tried to compute the strong coupling limit from
the Eden-Staudacher ran into some pathology as well: the result
obtained by Kotikov and Lipatov \cite{KL06} oscillates strongly, the
result obtained numerically in \cite{Benna-I} may not converge to a
straight line in $g$ as expected\footnote{We thank the authors of
\cite{Benna-I} for checking the behavior of their ES solution.} and
the numerical solution for $\sigma(u)$ in \cite{ES} does not seem to
converge to a value with definite normalization when $g\to \infty$.
We believe that all these pathologies are related to the separation of
the density $\rho(u)$ into $\rho_0(u)$ and $\sigma(u)$.
%This separation works if $K(u,u')\to 0$ vanishes sufficiently rapidly
%when $u\to \infty$ and for any value of $u'$.  This is not the case
%for the kernel without the dressing phase in the equation
%(\ref{Kss}), as it can be seen if we go to the $s$ variable
%\begin{equation} \Kpm(s,s')=\frac{1}{2\pi }\, \frac{1}{\cosh
%(s-s')}\to \frac{1}{2\pi }\, \frac{1}{\cosh (s')}\quad {\rm when}
%\quad s\to 0 \ (u\to \pm\infty)\;.  \end{equation}
Rather surprisingly, this pathology disappear when the dressing kernel
is taken into account, as we will show in the section 6.

\section{The dressing kernel}
\label{dk}
\setcounter{equation}{0}

\subsection{Integral representation}

Beisert, Eden and Staudacher proposed an expression for the dressing
factor in the Bethe ansatz equations.  This dressing factor translates
into a correction to the kernel in the integral equation.  Their
formula is given in Fourier transformed form; for our purposes an
expression in terms of variables $u$ is more suited.  Let us start
with the Fourier transform of the $su(1|1)$ kernel\footnote{Our
definition of the kernel differs by a minus sign from the one of
\cite{ES}.}
\begin{eqnarray}
 \hat K(t,t')=-\pi (1-\sgn tt') |t| e^{-(|t|+|t'|)\e}\;\sum_{n>0} n\;
 \frac{J_n(|t|)J_n( |t'|)}{|tt'|}\;.  \nonumber
\end{eqnarray}
which is related to the kernel of \cite{BES} by
\begin{equation}
\hat K(t,t')=-\frac{\pi}{2} (1-\sgn tt')|t| e^{-(|t|+|t'|)\e }\hat
K_m(|t|,|t'|) \,  
\end{equation}
 and can be broken into a symmetric and antisymmetric part 
\begin{eqnarray}
\hat K(t,t')&=&\hat K_+(t,t')+\hat K_-(t,t') \cr &=&-\frac{\pi}{2}
(1-\sgn tt')|t|e^{-(|t|+|t'|)\e } (K_0(|t|,|t'|)+K_1(|t|,|t'|))\;.
\nonumber
\end{eqnarray}
In the ES and BES equations, the kernel is written in the Fourier
variables defined only for positive values of $t$ and $t'$.  This is
made possible since the densities $\rho(u)$ and $\sigma(u)$ are
symmetric under $u\to -u$, and therefore under $t\to -t$, for the
states under consideration.  For generic states, this symmetry is not
present, and this is why it is desirable to have a non symmetric kernel
at hand.  The extension of $K_d(t,t')$ to negative values of $t$ and
$t'$ is ambiguous, and undoing the Fourier transform in the ``magic''
formula of \cite{BES} will be ambiguous as well.

The authors of \cite{Benna2} gave a prescription to to define
$K_d(u,u')$ by symmetrizing all the kernels on the second variable
$K(u,u')\to \half (K(u,u')+K(u,-u'))$.  We try to avoid symmetrizing
the kernels from the beginning, and give here a different
prescription, using instead that the full BES kernel is antisymmetric
with respect to $u_\pm \to u_\mp$ on each variable separately.  We
remind that the transformations $u\to -u$ and $u_\pm \to u_\mp$
translate in the following way on the variable $p$
\begin{eqnarray}
u\to -u \qquad &\Leftrightarrow& \qquad p\to 2\pi-p\\ \label{ptomp}
u_\pm \to u_\mp \qquad &\Leftrightarrow& \qquad p\to -p
\end{eqnarray}
while the variable $\beta$ stays unchanged.  Of course, these two
transformations act differently on the Fourier transform.

 It will be simpler to work with the non-symmetrized kernels and
 impose the antisymmetry under (\ref{ptomp}) at the end of the
 computation.  We define, for $t,t'>0$,
\begin{eqnarray}
\hat K_d(t,-t')&\equiv& -4\pi te^{-(t+t')\e}\hat K_c(t,t')\cr &=&-4\pi
te^{-(t+t')\e} \int_{0}^\infty dt'' \hat K_1(t,t'')\frac{t''}{e^{2\e
t''}-1}\hat K_0(t'',t') \nonumber \cr &=&-\frac{4}{\pi}
\int_{-\infty}^\infty dt'' \hat
K_-(t,-t'')\frac{e^{2\e|t''|}}{e^{2\e|t''|}-1}\hat K_+(t'',-t')
\;.\nonumber
\end{eqnarray}
For any values of $t,t'$, we can define the dressing kernel as a
convolution
\begin{equation}
\label{magic}
\hat K_d(t,t')= -\frac{4}{\pi} \int_{-\infty}^\infty dt'' \hat
K_-(t,t'')\frac{e^{2\e|t''|}}{e^{2\e|t''|}-1}\hat K_+(-t'',t')\; .
\end{equation}
The dressing kernel $\hat K_d(t,t')$, defined in this way, is
proportional to $1-{\rm sign}\; tt'$.  Note, however, that after
anti-symmetrization under (\ref{ptomp}) this property will not hold
anymore.  Going back to the $u$ variables we obtain
\begin{equation}
\label{repint}
K_d(u,u')=-8 \int_{-\infty}^\infty dv\; K_-(u,v)
K_+(v,u')-\frac{4}{\e} \int_{-\infty}^\infty dv\; dv'
K_-(u,v)h(v-v')\, \Theta _+(v',u')
\end{equation}
where 
\begin{equation}
h(u)=\frac{2\e}{2\pi} \int_{-\infty}^\infty dt
\;\frac{|t|e^{itu}}{e^{2\e|t|}-1}=\frac{2\e}{2\pi}
\left(\frac{1}{u^2}-\frac{(\pi/2\e)^2}{\sinh^2(\pi
u/2\e)}\right)
\end{equation}
and $\Theta_+(u,u')$ is the Fourier transform of $\hat K_+(t,t')/|t|$.
The function $h(u)$ becomes $\delta(u)$ as $\e$ approaches $0$.  Since
the kernels involved in the convolution in (\ref{repint}) are of order
$\e^0$, the dressing kernel will be of order $\e^{-1}$.  This formula
is very close to the one in \cite{Benna2}, except that we traded the
difficulty in evaluating the function $h(u)$ for the difficulty in
evaluating the phase $ \Theta_+(u,v)$.

Let us now write down the expressions of $ K_+(u,u')$ and $K_-(u,u')$,
which contain the odd, respectively even powers in the expansion of
the logarithm
\begin{eqnarray}
K_-(u,v)&=&\frac{1}{4\pi i} \partial_u
\ln{ (1-X_{+-})(1+X_{+-})\over (1-X_{-+})(1+X_{-+})}\;,
\no \\\no \\
K_+(u,v)&=&\frac{1}{4\pi i} \partial_u
 \ln {(1-X_{+-})(1+X_{-+})\over (1+X_{+-})(1-X_{-+})} 
  \, ,  
   \\\no \\
\Theta_+(u,v)&=&-\frac{1}{4\pi} \ln{ (1-X_{+-})
(1-X_{-+})\over(1+X_{+-}) (1+X_{-+}) } \;,
\la{auxkern}
\end{eqnarray}
with
\begin{equation}
X_{+-}=\frac{1}{x^+(u)x^-(v)}\;, \qquad X_{-+}=\frac{1}{x^-(u)x^+(v)}\;.
\end{equation}
A strategy to compute the expansion of $K_d$ in the strong coupling
limit is to compute the limit of the auxiliary kernels
(\ref{auxkern}), and then perform the integrals, as it was done in
\cite{Benna2}.  This procedure becomes rapidly quite involved, since
different pieces of the kernel contribute for different orders in
$\e$.

In the following, we give yet a different representation for the
dressing kernel, which may be more useful for the strong coupling
expansion.  The equation (\ref{magic}) can be written as
\begin{equation}
\hat K_d(t,t')= -\frac{4}{\pi}\sum_{n\geq0} \int_{-\infty}^\infty dt''
\hat K_-(t,t'')e^{-2n\e |t''|}\hat K_+(-t'',t')\, .
\end{equation}
This formula can be easily Fourier transformed back such that we get
\begin{equation}
K_d(u,u')=-8\sum_{n\geq1} \int_{-\infty}^\infty dv\; K_-^{1,n}(u,v)\;
K_+^{n,1}(v,u') \la{drek}
\end{equation}
with 
\begin{eqnarray}
K_\mp^{m,n}(u,v)&=&-\frac{1}{2\pi i} \partial_u
 \sum_{ l>0;\  ^{\rm even}_{\rm odd} } \frac{1}{l}
\Big((x^{(+m)}(u)x^{(-n)}(v))^{-l}-(x^{(-m)}(u)x^{(+n)}(v))^{-l}\Big)\; ,
\no \\
x^{(\pm n)}(u)&=&{u\pm in\e+\sqrt{(u\pm in\e)^2-1}}\;.
\end{eqnarray}
The variables $x^{(\pm n)}$ appeared in \cite{BHL}, where they play an
important role in defining the dressing phase.  They live on a torus
with modulus defined by
$$\frac{k'_n}{k_n}=n\e\;$$
and they are associated with the bound states of magnons \cite{Dorey}. 

The expression (\ref{drek}) for the dressing kernel involves a sum of
integrals along an infinite set of contours $\CC_{n\e} ^\infty$,
$n=1,2,3,\dots$, depicted in Fig.  5, defined in the same way as the
contour $\CC^\infty$, but with $\e$ replaced by $ n\e$ 
\eqn\opom{\CC_{n\e} ^\infty = \{x\in \IC\ \big|\ \( x - \bar x \) \(1-
1/ x\bar x \)= 4 i n\e , \quad \bar x x>1\}\, .  \la{cctrn} }
\vskip 25pt \centerline{   \epsfxsize=165pt \epsfbox{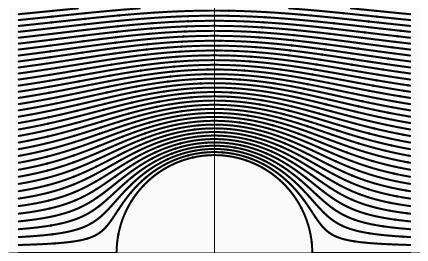} 
\hskip 1cm   \epsfxsize=150pt \epsfbox{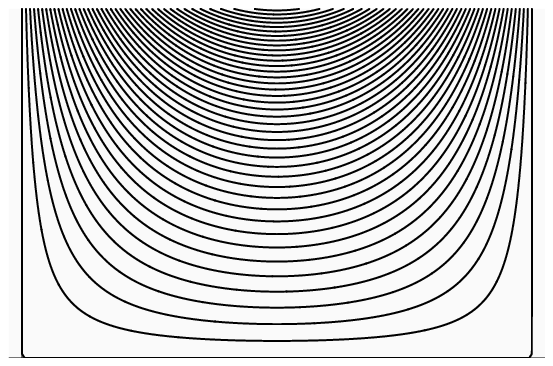} 
} 
\hskip 2.5cm -1 \hskip 1cm 0 \hskip 1cm 1\hskip 2.7cm 0 \hskip 4.5cm
$2\pi$

\bigskip
\begin{center}
{\small Fig.  5: The sequence of contours in the definition of the
dressing kernel \\ drown in the $x$-plane (left) and in the
$(p+i\b)$-plane (right).  }
\end{center}

\vskip 1cm

In the strong coupling limit, the expression (\ref{drek}) becomes
tractable, since we can evaluate the sum using the Euler-Maclaurin
formula
\eqn\McLaurin{
\sum_{n=1}^\infty f(n\e) ={1\over \e}  \int_0^\infty f(z)\, dz-\hf f(0) -
\sum_{k=1}^\infty    \e ^{2k-1} {B_{2k}\over (2k)!}   f^{(2k-1)}(0) .
}
We thus extend $n\e$ to the continuous
variable
\eqn\defz{ 
  {n\e} = z
}
and define the contour $\CC_z^\infty $ by
\eqn\defz{ \CC^\infty_z:\qquad \sin
{p\over 2} \sinh {\b \over 2} = z, \quad \b>0.  \la{defz} }
The integrals over $v$ and the variable $z$ become a double integral
over the domain $\CD$ in the complex $x$-plane spanned by the contours
$\CC^\infty_z$.  The domain in question is the the upper half plane
minus the unit semi-disk.  In terms of the complexified momentum $\b+
ip=2 \ln x$, the domain $\CD$ is the semi-infinite cylinder
$\CD=\{0\le p<2\pi, \b>0\}$.

The integral can be done without specifying a particular
parametrization of the contour $\CC_z^\infty$ if we express the
integrand as a wedge product of differential forms.  We first write
the basic kernel as a 1-form
\eqn\defK{ K(x,x_1) = - {dp\over 4 \pi }
+ {1\over 4 \pi } { \sinh{\beta +\beta _1\over 2} \, dp-\sin
{p-p_1\over 2} d\beta \over \cosh {\beta+\b_1\over 2} -
\cos{p-p_1\over 2} } .  \la{defKtd}}
Then the ``even'' kernel $K_+ $ is a one-form in the basis of $dp$ and
$d\b$,
\begin{eqnarray}
K_+(x,x_2)&=&\frac{1}{2\pi}\frac{ \sinh{\beta+\b_2\over 2}\cos{p-
p_2\over 2} \, dp - \sin{p- p_2\over 2}\cosh{\beta+\b_2\over 2} \,
d\beta} {\cosh(\beta+\b_2)-\cos(p- p_2)} \, ,
    \end{eqnarray}
while the ``odd'' kernel $K_-$ is a zero-form  in this basis,
\begin{eqnarray}
 K_-(x_1,x)
 &=&-\frac{1}{4\pi}{dp_1} +
  \frac{1}{4\pi}\frac{d p_1\sinh(\beta+\b_1)-d\b_1 \sin(p_1- p)}
 {\cosh(\beta+\b_1)-\cos(p_1- p)} \, .
% &=& \frac{1}{4\pi}\frac{d p_1[\cos(p_1-p) - e^{-|\beta+\b_1|}]
% -d\b_1 \sin(p_1- p)} {\cosh(\beta+\b_1)-\cos(p_1- p)} \;,\\
\end{eqnarray}
Also, for the differential $dz$ we have from (\ref{defz})
 \eqn\defdz{ {dz } = \half \sinh(\b/2)\cos(p/2) \, dp+\half
 \sin(p/2)\cosh(\b/2) \, d\b\, .  \la{defdz}}
Then the leading term in the strong coupling limit $\e\to 0$ is given
by
 \eqn\drLnn{ K_d^0(x_1,x_2) =-{8\over \e} \int_0^\infty \!\!\!  dz
\int _{x\in \CC_z} \!\!\!  K_- \; K_+ \ \equiv \ -{8\over \e}
\int_{\CD} \; K_- \; dz\wedge K_+ .
\label{Lnn}}
The evaluation of this integral (the calculation is sketched in
Appendix B) leads to the following expression for $K_d^0$
\begin{eqnarray} \label{kdo}
   K_d^0(x_1,x_2)&=&\frac 1{4\pi \e}\,
   \partial_u\left[-\tilde\chi(x_1^-,x_2^+)+\tilde\chi(x_2^+,x_1^-)
   +\tilde\chi(x_1^-,-x_2^+)-\tilde\chi(-x_2^+,x_1^-)
   +{\rm c.c.}\right],\no \\
   \tilde\chi(x,y)&=&-\(x+\frac 1x\)\log\(y-\frac 1x\) .
\end{eqnarray}
This expression must be antisymmetrized with respect to
$x_2^{\pm}\rightarrow x_2^{\mp}$ (or, equivalently, $p_2\rightarrow
-p_2$, $\beta_2\rightarrow \beta_2$).  After antisymmetrization the
result coincides\footnote{It is useful to notice that we can
substitute $\chi_0$ with $\tilde\chi$ to calculate AFS term in
(\ref{tafs}).} with the AFS kernel $K^{AFS}(u,v)$, symmetrized with
respect to $v\rightarrow -v$.

\subsection{ The strong coupling limit from the AFS kernel}
 
To compute the leading order(s) of the dressing kernel, we do not
really need the integral representation derived in the previous
subsection.  It is more straightforward to use the strong coupling
representation of the dressing phase $\th_{12} = \th(u_1, u_2)$ worked
out in \cite{BHL,BES}.  The latter can be cast in the form
\begin{eqnarray}\la{tafs}
\theta_{12}&=&\frac{1}{2\e}\[\chi^{--}_{12}
-\chi^{+-}_{12}-\chi^{-+}_{12}+\chi^{++}_{12}-(1\leftrightarrow
2)\]\, ,\\
\nonumber
\chi_{12}^{rs}&=&\chi\(x_1^r,x_2^s\)\;, \quad r,s=\pm\, ,
\end{eqnarray}
where the function $\chi$ can be expanded in powers of the inverse
coupling constant
\begin{equation}
\chi=\sum_{n\geq 0} \chi_n (2\e)^n\, .
\la{chinn}
 \end{equation}
To compute the leading term in the anomalous dimension, the AFS term
\cite{AFS} is sufficient,
\begin{equation}
\chi_0(x,y)=-\frac{xy-1}{y}\ln\frac{xy-1}{xy}\, ,
 \end{equation}
while for the next  correction we will need the Hern\'andez-L\'opez 
\cite{HL,AF} term 
\begin{eqnarray}
&&\chi_1(x,y)=\frac{1}{\pi}\Big[\ln\frac{y-1}{y+1}\ln\frac{x-1/y}{x-y}\\
\nonumber &&+\Li_2\frac{\sqrt{y}-1/\sqrt{y}}{\sqrt{y}-\sqrt{x}}
-\Li_2\frac{\sqrt{y}+1/\sqrt{y}}{\sqrt{y}-\sqrt{x}}+\Li_2\
\frac{\sqrt{y}-1/\sqrt{y}}{\sqrt{y}+\sqrt{x}}-
\Li_2\frac{\sqrt{y}+1/\sqrt{y}}{\sqrt{y}+\sqrt{x}}\Big]\, .
 \end{eqnarray}
In the following, we give the first non-zero order of the dressing AFS
phase.  The kernel can be simply obtained by taking the derivative of
the phase divided by $2\pi$, with the right sign.
We find,  for the sectors $``<"$ and $``>"$,
 \begin{eqnarray}
 \tpp(u_1,u_2)&=& \frac{2\e}{\sinh \b_1/2 \sinh \b_2/2}\;\frac{\sinh
 (\b_1-\b_2)/4} {\sinh (\b_1+\b_2)/4}-2\vppp(u_1,u_2) +\CO(\e^2) \no
\, , \\\nonumber \tpm(u_1,u_2)&=&-\frac{2\,\sin p_2/2}{\sinh \b_1/2}
 -2\vppm(u_1,u_2) +{\cal O}(\e )\, ,\\\nonumber \tmp (u_1,u_2) &=&\ \
 \frac{2\,\sin p_1/2}{\sinh \b_2/2} -2\vpmp(u_1,u_2) +{\cal O}(\e)\,  ,\\
 \tmm (u_1,u_2) &=& \frac{1}{\e}\Big[(\cos p_1/2-\cos p_2/2)
 \ln \frac{\sin^2(p_1-p_2)/4}{\sin^2(p_1+p_2)/4}
 \Big]+{\cal O}(1)\, . \la{leadingkerda}
  \end{eqnarray}
 \medskip
 We remind thet $\vp$ is the $su(1|1)$ phase.

In the sectors $``+''$ and $``-''$, where $\b$ and $p$ (or $2\pi - p$)
are of order $\sqrt{\e}$, we introduce the parameter $\a$, which is of
order one, as follows
\begin{eqnarray}
 p_i&=& \quad 2\sqrt{\e}\, \a_i\, , \qquad \b_i =2 \sqrt{\e}\,
 \a_i^{-1} \ \ {\rm with } \ \ \a_i>0 \qquad {\rm in\ the\ region}
 ``+"\, ,\no \\
 p_i&=& 2\pi + 2\sqrt{\e}\, \a_i, \ \ \b_i = - 2\sqrt{\e}\, \a_i^{-1}
 \ \ {\rm with } \ \ \a_i<0 \qquad {\rm in\ the\ region} ``-" .  \no
 \end{eqnarray}
 Then we have
  
\begin{eqnarray}
  \nonumber \theta^{\pm\pm}(\a_1,\a_2)&=&\pm
  \frac{1}{2}\Big[(\a_1^{-2}-\a_1^2-\a^{-2}_2+\a_2^2) \ln
  \frac{(\aa_1+\aa_2)^2+(\a_1-\a_2)^2}{(\aa_1+\aa_2)^2+(\a_1+\a_2)^2}
\, ,  \\ \no && \ -
  4i\ln\frac{\aa_1+\aa_2-i(\a_1-\a_2)}{\aa_1+\aa_2+i(\a_1-\a_2)}\Big]
  +{\cal O}(\sqrt{\e})\, , \\\nonumber 
  \theta^{\pm\mp}(\a_1,\a_2)&=&\pm
  2\; \a_1\a_2+{\cal O}(\sqrt{\e})\\ \nonumber
  \theta^{\pm>}(\a_1,u_2)&=&\pm{2\sqrt{\e}}\; \frac{\a_1}{\sinh \b_2
  /2} +{\cal O}(\e)\, ,\\
 \nonumber
  \theta^{\pm<}(\a_1,u_2)&=&-\frac{2}{\sqrt{\e}} \;\a_1
  \sin p_2/2+{\cal O}(1) \, ,\\ \nonumber
  \theta^{>\pm}(u_1,\a_2)&=&\mp{2\sqrt{\e}}\
  \frac{\a_2}{\sinh \b_1 /2}+{\cal O}(\e)\, ,\\
 \theta^{<\pm}(u_1,\a_2)&=&\frac{2}{\sqrt{\e}}\;\a_2\sin p_1/2+{\cal
 O}(1)\, . \la{leadingkerdb}\end{eqnarray}
 %
% where $p_i=2\sqrt{\e}\a_i$ with $\a_i>0$ in the region ``+'' and
% $p_i=2\pi + 2\sqrt{\e}\a_i$ with $\a_i<0$ in the region ``-'' and
% $\b_i=\pm 2\sqrt{\e}\aa_i$.
%
In (\ref{leadingkerda}) and (\ref{leadingkerdb}) we omitted the signs
coming from $u$ being in $u>1$ and $u<-1$.  A mnemonic rule to
reproduce these signs is to take ${\rm sign}\, \b_i = {\rm sign}\,
u_i$.  In the hyperbolic parametrization with the variable $s$, which
we give below, the signs are automatically taken into account
 \begin{eqnarray}
 \label{AFSker}
 \tpp(s_1,s_2)  &=&2\e \sinh s \sinh s'
 \tanh\frac{s-s'}{2}-2\vppp(s_1,s_2) +\CO(\e^2)\\\nonumber
 \tpm(s_1,s_2) &=&- \frac{2\sinh s_1}{\cosh s_2}
 -2\vppm(s_1,s_2) +\CO(\e) \\\nonumber
 \tmp  (s_1,s_2)&=& \frac{2\sinh s_2}{\cosh s_1} 
 -2\vpmp(s_1,s_2)+\CO(\e) \\\nonumber
 \tmm (s_1,s_2)&=&\frac{2}{ \e}\(\frac{\sinh(s_1-s_2)}{\cosh s_1 \cosh s_2}
 \ln |\tanh {s_1-s_2\over 2}|
\)+{\cal O}(1)\, .\no
%\\
% &=& {1\over 2\pi \e} \(-{\cosh s} \ \log |\tanh{s-s'\over 2}| -
% {1\over \cosh s'} \) +{\cal O}(1)\nonumber
  \end{eqnarray}
  
   \medskip
   
In the regimes where one of the variables is in the region $``+"$ or
$``-"$, we put $\a_i=\pm e^{\pm s_i}$ 
\begin{eqnarray}
 \label{thpm}
 \theta^{_{\pm\pm}}(s_1,s_2)&=&\mp  \Big[(\sinh 2s_1-\sinh 2s_2)\ln
 \frac{1+e^{\pm2(s_1+s_2)}\tanh^2\frac{s_1-s_2}{2}} {1+e^{\pm2(s_1+s_2)}}\\
 \no && +4\arctan \(e^{\pm(s_1+s_2)}\tanh\frac{s_1-s_2}{2}\)\Big] +{\cal
 O}(\sqrt{\e})\,  ,\\\nonumber
 \theta^{_{\pm\mp}}(s_1,s_2)&=&\mp2e^{\pm(s_1-s_2)}+{\cal
 O}(\sqrt{\e})\, ,\\ \nonumber
   \end{eqnarray}
  \begin{eqnarray}
  \theta^{_{\pm>}}(s_1,s_2)&=&2\sqrt{\e}\;e^{\pm s_1}\sinh s_2+{\cal
  O}(\e)\, , \\\nonumber \theta^{_{\pm<}}(s_1,s_2)&=&\mp
  \frac{2}{\sqrt{\e}}\; \frac{e^{\pm s_1}}{\cosh s_2}+{\cal O}(1)\, ,\\
  \nonumber \theta^{_{>\pm}}(s_1,s_2)&=&-2\sqrt{\e}\;e^{\pm s_2}\sinh
  s_1+{\cal O}(\e)\, ,\\ \no \theta^{_{<\pm}}(s_1,s_2)&=&\pm
  \frac{2}{\sqrt{\e}}\;\frac{e^{\pm s_2}}{\cosh s_1}+{\cal O}(1)\, .
 \end{eqnarray}
 It is important to note that, in what concerns $\theta^{\pm\pm}$, the
 equation (\ref{thpm}) is not the correct answer, since in this region
 all terms $\chi_n$ in (\ref{chinn}) contribute to the $\CO(1)$ order.
 This was pointed out by Maldacena and Swanson
 \cite{Maldacena-Swanson}, who derived an integral representation for
 the scattering phase in this region.
   
  It is interesting to note
 that $\theta^{\pm\pm}$ and $\theta^{\pm\mp}$ in eq.  (\ref{thpm})
 depend naturally on the rotated coordinates
$$s=s_1+s_2\;, \qquad   \tilde s =s_1-s_2\;,$$
and that for large values of $s$, positive or negative, we have
$$ \theta^{_{++}}(s_1,s_2)\simeq 2 e^{s} 
(\tanh \frac{\tilde s}{2}-1-\sinh \tilde s)\;.
$$

\section{Solving the integral equations in the strong coupling limit
with the dressing kernel}
\label{sclsold}
\setcounter{equation}{0}

Once we have obtained the leading term of the dressed kernels, we are
able to compute the leading contribution to the energy of the states
we are interested in.  In the $su(1|1)$ and $su(2)$ sectors, we do not
need to obtain the functional form for the density explicitly.  This
is a fortunate situation, since in the $(\pm\pm)$ regime, where most
of the roots are concentrated, we do not know the kernel in a closed
form.  The computation is very similar to the one performed in
\cite{AFS}, where the $\lambda^{1/4}$ behavior for the anomalous
dimensions was first derived.

\subsection{The $su(1|1)$ dressed case}

The results of the numerical and analytical computation by Beccaria
{\it et al.} \cite{Beccaria-I, Beccaria-II} for the energy of the
highest excited state in the $su(1|1)$ sector can be also derived from
the integral equation.  As it was shown in \cite{Beccaria-II}, the
same results can be obtained using the light-cone Bethe ansatz
equations \cite{LCBA}, first derived in the $su(1|1)$ sector by
Arutynov and Frolov \cite{AFsu11}.  In this sector, the total kernel
is
\begin{equation}
\CK =K+K_d\;.
\end{equation}
First, we can show that the densities $\rho_<$ and $\rho_>$ are
subdominant, while the main contribution comes from $\rho_\pm$.  The
density in the "giant magnons" region is given by the equation
   \begin{eqnarray}
   \label{rhoin}
\rho^<=\frac{1}{2\pi} \frac{dp}{ds} + \CK^{<<}\rho^<+
\CK^{<>}\rho^>+\sum_\pm \CK^{<\pm}\rho^\pm \,.
\end{eqnarray}
By symmetry, we have $\rho_+(\a)=\rho_-(-\a)$ and since
$$\int_0^\infty \a\rho_+(\a)d\a +\int_{-\infty}^0 \a\rho_-(\a)d\a=0
\, , $$
  we conclude that the $\pm$ terms in (\ref{rhoin})
cancel 
\begin{equation}
\rho^<=\frac{1}{2\pi} \frac{dp}{ds} + \CK^{<<}\rho^<+
\CK^{<>}\rho^>+{\cal O}(1)\;.
\end{equation}
All the terms except $\CK^{<<}\rho^<$ are of the order $1$.  Since
$\CK^{<<}=\CO(1/\e)$ we deduce that $\rho_<(s)={\cal O}(\e)$.
Similarly, we can write
    \begin{eqnarray}
\rho^>=\frac{1}{2\pi} \frac{dp}{ds}
+\CK^{><}\rho^<+
\CK^{>>}\rho^>+\sum_\pm \CK^{>\pm}\rho^\pm\;.
\end{eqnarray}
Again the leading term in the sum vanishes and all the terms in the
r.h.s. are of the order $\e$, so that $\rho_>(s)$ is also of order at
most $\e$.  Therefore the roots of the Bethe equations must
concentrate in the regions $\pm$, or "near-flat space" region.  In
this region, we have
\begin{eqnarray}
\rho^\pm=\frac{1}{2\pi } \frac{dp}{d\a} + \CK^{\pm<}\rho^<+
\CK^{\pm>}\rho^>(s')+ \CK^{\pm\pm}\rho^\pm+  \CK^{\pm\mp} \rho^\mp \, .
\end{eqnarray}
At the leading order, we are left with
   \begin{eqnarray}
\rho^\pm(\a)= \int d\a' \CK^{\pm\pm}(\a,\a')\rho^\pm(\a')+ \int d\a'
\CK^{\pm\mp}(\a,\a') \rho^\mp(\a')
\end{eqnarray}
or, by specializing to the sign $+$ and by integrating over $\a$,
\begin{eqnarray}\label{kp}
k^+(\a)=-\frac{1}{2\pi} \int\limits_0^\infty d\a' \,\theta^{++}_{\rm
tot}(\a,\a')\rho^+(\a')-\frac{1}{2\pi}\int\limits_{-\infty}^0 d\a'
\,\theta^{+-}_{\rm tot}(\a,\a') \rho^-(\a')\;,
\end{eqnarray}
where $k^+(\alpha)$ is the counting function, $k^+(\a)\equiv\int_0^\a
d\a\, \rho^+(\a)$.  $\theta^{+\pm}_{\rm tot}$ is defined as
$\theta^{+\pm}_{\rm tot}= - 2\pi\int_0^\a d\a\, \CK^{+\pm}$ and at
leading order is equal
%\footnote{Althougth the true expression for $\theta^{++}$ is 
%different \cite{Maldacena-Swanson}, we need only the antisymmetric 
%property of $\theta^{++}$, which is always valid.} 
to the dressing part $\theta^{+\pm}$. By multiplication with $\rho(\a)$
and integration we get
\begin{eqnarray}
   \label{xxp}
\int_0^\infty d\a \; k^+(\a)\rho^+(\a) &=&-\frac{1}{2\pi}\int_0^\infty
d\a \int_0^\infty d\a' \theta^{++}(\a,\a^\prime)\rho^+(\a)\rho^+(\a') \\
\nonumber &-&\frac{1}{2\pi}\int_0^\infty d\a \int_{-\infty}^0 d\a'
\theta^{+-}(\a,\a')\rho^+(\a) \rho^-(\a')\,.
\end{eqnarray}
Here we used that $\theta^{+-}(0,\a')=\theta^{++}(0,\a')=0$.  Let us
remind that $\alpha=0$ corresponds to $u=\infty$.  Due to the
anti-symmetry of the phase $\theta^{++}$, the first integral in the
r.h.s. of (\ref{xxp}) vanishes, while the l.h.s. is equal to
\eqn\lhsrho{\label{kzeromax} \int_0^\infty d\a \; k^+(\a)\rho^+(\a) = 
\int_0^{1/2}
dk^+ \; k^+=\frac{1}{8}\;.
}
Finally, we use that $\theta^{+-}(\a,\a')=2\a\a'$ and the symmetry
$\rho^+(\a)=\rho^-(-\a)$ to cast the equation (\ref{xxp}) in the form
\begin{equation}
\int_0^\infty d\a \; \a\; \rho^+(\a) =\sqrt{\frac{\pi}{8}}\;.
\end{equation}
This is all we need to compute the leading order in the energy
   \begin{eqnarray}
E_{su(1|1)}^d=4gL\int e^{-\beta/2}\sin p/2\; \rho(p)\; dp\simeq
4\sqrt{g}L\int_0^\infty d\a\; \a \; \rho^+(\a) =\sqrt{2\pi g}L\;.
\end{eqnarray}
It is remarkable that the energy of this state is of the order
$\lambda^{1/4}$.  It would be interesting to know if there are states
in the $su(1|1)$ sector which have greater energy.  This would be
possible for states where the fraction of giant magnons is larger than
$1/\sqrt{g}$.  Presumably, the strong repulsion among the giant
magnons will prevent this phenomenon to happen.

\subsection{The su(2) dressed case}

The computation of the leading order in the energy for $su(2)$ sector
goes similarly to that for $su(1|1)$, with two differences.  First,
there is a non-trivial contribution from the $su(2)$ kernel to the
leading order of $\theta^{++}_{{\rm tot}}$
\begin{equation}
	\theta_{su(2)}(u,u')=2\pi \int\limits_{\infty}^u dv\;
	K_{su(2)}(v,u')=2 \int_{u}^\infty dv\;
	\frac{2\e}{(v-u^\prime)^2+4\e^2}=\pi+\theta_{asym}(u,u')\,.
\end{equation}
Therefore, the first term in the r.h.s. of the equation (\ref{xxp})
does not cancel anymore and we obtain
\begin{eqnarray}
   \label{xxpsu2}
\int_0^\infty d\a \; k^+(\a)\rho^+(\a) =-\frac{1}{2}\(\int_0^\infty
d\a\; \rho^+(\a)\)^2 
 +\frac{1}{\pi}\(\int_0^\infty d\a  \;
\a \;\rho^+(\a)\)^2\,.
\end{eqnarray}
Second, the maximum of counting function is 1/4 instead of 1/2,
because the total number of magnons is $M=L/2$, and
$$\int_0^\infty d\a \;\rho^+(\a)=\int_0^{1/4} dk^+=\frac{1}{4}\;.$$  
From  equation (\ref{xxpsu2}) we obtain
\begin{eqnarray}
	 \int\limits_0^\infty d\a \; \a\; \rho^+(\a)=\sqrt{\frac{\pi}{16}}\;,
\end{eqnarray}
which reproduces\footnote{We thank M. Beccaria for an email exchange
which helped us to remedy a discrepancy with their results in the
first version of this paper.} the value of the energy obtained by
Beccaria {\it et al.} \cite{Beccaria-I, Beccaria-II}
\begin{eqnarray}
	E_{su(2)}^d=\sqrt{\pi g}L+{\cal O}(g^0)\;.
\end{eqnarray}

\subsection{The $sl(2)$ dressed case}

For the $sl(2)$ case, the total kernel with the $su(2)$ piece
subtracted, as defined in (\ref{totker}), is equal to
\begin{eqnarray}
\CK(u,u')=2K(u,u') +K_d(u,u')\equiv {1\over 2\pi} \p_u\phi(u,u')\, .
\end{eqnarray}
The BES equation for this sector takes the simplest form when written
for the shifted density function \eqn\defbarsig{ \la{defbarsig}
\srb(u)\equiv\sigma(u)- \frac{2\e}{\pi}\, , } which, up to a negative
factor, equals the total density at the leading order in $S$, see
(\ref{rhosig}).  The BES equation reads
\begin{eqnarray}
\label{eqnsl2d}
&&\srb(u)= \frac{1}{\pi}\int_{-\infty}^\infty
du'\;\frac{2\e\;\srb(u')}{(u-u')^2+4\e^2}\; +\int du'\; \CK(u,u')\;
\bar\sigma(u')\; .
\end{eqnarray}

We are interested in expanding equation (\ref{eqnsl2d}) in
powers of $\e$.   At small $\e$
\be \frac{2\e}{\pi}\pint_{-\infty}^\infty
du'\;\left(\frac{\srb(u')}{(u-u')^2}+\CO(\e^2)\right)\; +\int du'\;
\CK(u,u')\; \bar\sigma(u')=0\;.
\ee
This equation splits into several equations, which couple the
different regions in $u$.  For instance, if $|u|<1$ we have
\begin{eqnarray}
\label{eqin}
\nonumber
&&\frac{2\e}{\pi} \(\pint_{-1}^1 du'\;\frac{\srb^{_<}(u')}{(u-u')^2}+
\int_{>} du'\;\frac{\srb^{_>}(u')}{(u-u')^2}+\CO(\e^2)\)
+
\int_{-1}^1 du' \;
\CKmm(u,u')\;  \srb^{_<}(u')\\  &&+\int_{^>} du'\;
\CKmp(u,u')\srb^{_>}(u')+\sum_\pm\int d\alpha\;
\CK^{_<\pm}(u,\alpha)\; \srb^{\pm}(\alpha)=0\;.
\end{eqnarray}
%%
%while for $|u|>1$ we obtain
%%
%\begin{eqnarray} \label{eqout} &&\srb^{_>}(u)-\frac{1}{2\pi}
%\int_{^>}du'\;\frac{2\e\;\srb^{_>}(u')}{(u-u')^2+4\e^2} = \int_{^>}
%du' \; \CKpp(u,u')\;\bar\srb^{_>}(u')\\ \nonumber &&+\int_{-1}^1
%du'\; \CKpm(u,u')\bar\srb^{_<}(u')+\sum_\pm\int d\alpha\;
%\CK^{_>\pm}(u,\alpha)\;\bar\srb^{\pm}(\alpha)\;, \end{eqnarray}
%%
%with $\bar \srb(u)$ the shifted density $$\bar \srb(u)=
%\srb(u)-\frac{2\e}{\pi}\;.$$ A third equation can be written for
%$u=\pm1\[1-\e(\a^2-\a^{-2})/2\]$; we are not going to use it there,
%but it will be important for determining the density
%$\srb^\pm(\alpha)$.

Equation (\ref{eqnsl2d}) was analyzed in the first two orders in $\e$
by the authors of \cite{Benna2}.  They did not consider the regions
around $u=\pm1$, which can contain a fraction of the roots.  Equation
(\ref{eqnsl2d}) suggests that the $\e$-expansion of the density starts
at order $\e$.  The structure of the leading terms
(\ref{leadingkerdb}) in the expansion of the dressing kernel, where
half-integer powers of $\e$ appear, suggests that the density and the
anomalous dimension may involve a half-integer powers of $\e$ as well.
However, these corrections seem to appear in higher orders in the
anomalous dimension, and this may explain why they have not been seen
yet.\footnote{As far we can see, the first non-zero order for
$\sigma^\pm(\alpha)$ is at most $\e^{2}$, and there may be corrections
of order $\e^{5/2}$ to $\sigma^{_<}(u)$ and of order $\e^{3/2}$ for
$\srb ^{_>}(u)$.  The last of these corrections would induce a term of
order $g^{1/2}$ in the anomalous dimension.  To decide whether they
are here or not we have to go to higher orders in the expansion of the
kernel.} In the following, we are going to ignore the corrections
coming from the vicinity of the points $u=\pm1$ and solve
perturbatively the equation (\ref{eqnsl2d}), by considering the
decomposition of the density and the kernel in integer powers of $\e$
\begin{eqnarray}
\srb(u) &=&   \sigma(u)-\frac{2\e}{\pi}= \e
\srb_1(u)+\e^2\sr_2(u)+   \e^3  \sr_3 (u)+\ldots\;, \no \\ \no\\
\CK(u,u')&=&\e^{-1}\CK_{-1}(u,u')+\CK_0(u,u')+\e\CK_1(u,u')+\ldots\;.
\end{eqnarray}
%
%\bigskip

\noindent
{\bf $\bullet$ order $\e^{0}$}

\medskip
\noindent
At  the leading order in $\e$   equation (\ref{eqin}) reads, in
simplified notations,
$$\CK_{-1}\srb_1 =0.
$$
To this order only the $<<$ sector contributes
\begin{equation}
\int_{-1}^1 du' \; \CKmm_{-1}(u,u')\; \srb^{_<}_1(u')=0\;.
\end{equation}
It was proven in \cite{Benna2} that the kernel $\CKmm_{-1}$ is
non-degenerate, therefore we have $ \srb ^{_<}_1(u)=0$, or
equivalently $\sr _1=2/\pi$.  \bigskip

\noindent
{\bf $\bullet$ order $\e^{1}$}

\medskip
\noindent
 The $O(\e)$ term in the expansion of   equation (\ref{eqin}) 
 relates $\srb _1$ and $\sr _2$
\begin{eqnarray}
\CKmm_{-1}\;\sr ^{_<}_2+
\CKmp_0\srb ^{_>}_1
=0\; .
\end{eqnarray}
The kernel in the  second term  is  antisymmetric in $u'\to -u'$
\begin{eqnarray}
\CKmp_0(u,u')=- \frac{1}{\pi}\p_u\frac{\sqrt{1-u^2}}{\sqrt{u'^2-1}}\;
{\rm sgn}\, u'\;,
\end{eqnarray}
so that it vanishes upon integration against the symmetric function
$\srb _1^{_>}(u)$.  As pointed out in \cite{Benna2}, the subleading
density $\sr ^{_<}_2(u)$ vanishes as well.  It is worth noticing that
$\CKmp_0(u,u')$ is the sum of two terms, one coming from the $su(1|1)$
kernel and the other from the dressing phase,
$\CKmp_0(u,u')=2\Kmp_0(u,u') +\Kmp_{d,0}(u,u')$.
%If the relative factor of the two contribution was different, the
%structure of the solution would be different: either both $\sr
%^{_<}_2(u)$ and $\srb _1^{_>}(u)$ will be different from zero, or
%they will be both zero.  In the second situation we will be
%confronted again with the problem in the Eden-Staudacher equation.

 \bigskip

\noindent
{\bf $\bullet$ order $\e^{2}$}

\medskip

\noindent
At this order we have  an  equation 
 relating  $\srb _1$,  $\sr _2$ and $\sr _3$ 
$$
{2\over \pi} \pint _{-\infty}^\infty d u' {\srb_1(u') \over (u-u')^2}
+ \CK_{-1} \sr_3 + \CK_0 \sr _2+\CK_1\srb_1 =0\, .
$$

\bigskip
\noindent
$\circ$ The interval $|u|>1$

\bigskip
\noindent
Consider first the region $|u|>1$, where the equation takes the form
\begin{equation}
\label{eqdet}
\frac{2}{\pi}\pint _{>} du' \frac{ \srb ^{>}_1(u')}{(u-u')^2}+\int
_{>} du' \;\CKpp_1(u,u')\; \srb ^{>}_1(u') =0\, ,
\end{equation}
%
% label of (5.115)
with the kernel $\CKpp_1(u,u') \equiv\frac{1}{2\pi}\p_u \phi_1 $
given by (\ref{AFSker} ).  Equation (\ref{eqdet}) is a total
derivative, so we first integrate it to
\begin{equation}
\pint_{>} du' \srb ^{_>}_1(u') \( - {2\over\pi} \frac{1}{u-u'}+
{1\over 2\pi} \phi_1 (u, u') \)= 0 \, .\la{inteqa}
\end{equation}
As the phase is anti-symmetric, there is no integration constant.

At this point, we find it more useful to switch to the variable $s$,
defined by $u=\coth s$,
% We have \eqal\skern{ \phi_1(u,u') &=& 2\sinh s \sinh s'
% \tanh\frac{s-s'}{2}\; , \no\cr - \frac{4}{u-u'}+ \phi_1 (u, u')&=& 2
% \sinh s \sinh s' \, \coth{s-s'\over 2} \la{skern} }
%%
%%
and (\ref{inteqa}) becomes
\begin{equation}
{1\over \pi} \pint _{-\infty}^\infty ds'\, \srb ^{_>}_1(s') \sinh s'
\, \coth{s-s'\over 2}\, = 0 \, .  \la{inteqb}
\end{equation}
Now we can reformulate the integral equation (\ref{inteqa}) as a
Riemann boundary value problem.  We first rewrite (\ref{inteqb}) in
terms of the normalizable density $\sr _1 = \srb_1+ 2/\pi$ 
% Evaluating  the  integral
%%
%\eqn\integrvp{ \int_{-\infty}^\infty \phi_1 (s,s') \; |u'(s')| d s'
%%=\sinh s \int_{-\infty}^\infty d s' { \tanh\frac{s-s'}{2} -
%%\tanh\frac{s+s'}{2} \over \sinh s' }= - 2 \sinh s
%%\int_{-\infty}^\infty {d s' \over \cosh s +\cosh s' } =
% =-{4s } }
% (here we symmetrized the phase to avoid the ambiguity at
% $s'=0$), we write
%
\begin{equation}{1\over \pi}
   \pint _{-\infty}^\infty ds'\, \sr ^{_>}_1(s') \sinh s' \,
   \coth{s-s'\over 2}\, = {4s\over \pi^2 \sinh s} \, .  \la{intsr}
\end{equation}
Then, after the redefinition
\eqn\defor{ \o=e^{s} , \quad r(\o) = - r(1/\o) = \sr
^{>}_1(s)\, \sinh s\, , \la{defor} }
the l.h.s. of (\ref{intsr}) takes the form of a Cauchy integral 
\begin{equation}
\int _0^{\infty} {d\o' \over \o'}  r(\o') + 2 \pint _0^{\infty}
{d\o' } {  r(\o')\over \o-\o'} ={8\over \pi } \ {\ln\o \over \o -
\o^{-1} },\qquad (\o>0) .
\la{singin}
\end{equation}
The first term on the l.h.s. is actually zero due to the anti-symmetry
of $r(\o)$.  

Equation (\ref{singin}) can be formulated in terms of 
a boundary condition for  the  resolvent
\eqn\defG{ R(\o) =R(1/\o) = \int _0^{\infty}{d\o' } \ { r(\o')\over
\o-\o'} \, ,  }
which has a cut on the positive axis, namely,  
\eqal\diffeq{ R(\o+i0)+R(\o-i0) &= & {8\over \pi } \ {\ln \o \over \o
- \o^{-1}}\; \qquad (\o>0)\, .
 \la{diffeq} }
The most general solution of (\ref{diffeq}) 
with the symmetry $R(\o)=R(1/\o)$ is of the form
\eqn\gensR{ R(\o)= {4\over \pi } \, {\ln (-\o) + \sqrt{-\o} \, Q(\o) -
\sqrt{-1/\o}\, Q(1/\o) \over \o- 1/\o}\,  ,
}
where $Q(\o)$ is a rational function.  The latter is determined by the
analytical properties of the resolvent (no poles outside the positive
real axis) and the requirement that the density $\sr_1$ is
normalizable
$$
\int_> du\, \sr_1(u) = \int_{-\infty} ^\infty ds \,\sr_1^>(s) =
\int_0^\infty d\o \,{ 2 r(\o)\over \o^2-1} = {\rm finite}.
$$
The only solution with these properties corresponds to $Q=-1/2$:
\eqal\solutionR{ R(\o)&=& {4\over \pi } \, {\ln (-\o) - \hf\(
\sqrt{-\o}\, - \sqrt{-1/\o}\) \over \o- 1/\o}\; , }
and its discontinuity along the positive real axis,
\eqn\solr{ r(\o) ={ R(\o+i0) -R(\o-i0) \over 2\pi i } = {2\over \pi }
\({2\over \o- 1/\o} - {\sqrt{\o}\over \o-1}\)\; , } reproduces the
solution found by Alday {\it et al} \cite{Benna2}:
    \eqn\Kleb{ \sr_<(u) = \e \sr _1 ^< = {1\over 2 \pi g}, \quad
    \sr_>(u)= \e \sr_1^>(u) = {1\over 2 \pi g} \(1- \cosh {s\over 2}\).
    }
The total integral of the density,
\eqn\intdens{ \int _{-\infty}^\infty ds\, \( \sr^{_<}(s)+
\sr^{_>}(s)\) = {1\over \pi g} + {\pi-4\over 4\pi g} = {1\over 4g}\;,
}
reproduces the correct leading behavior for the twist-two anomalous
dimension \eqn\intdensi{ f(g)=4g+\ldots\;.  }

\bigskip
\noindent
$\circ$ The interval $|u|<1$

\bigskip \noindent Let us assume for the moment that the corrections
from the vicinity of the points $u=\pm 1$ can be neglected.  Then, in
the region $|u|<1$, the equation will reduce to
\be \frac{2}{\pi}\int_{|u'|>1} du'\;\frac{\srb
_1^>(u')}{(u-u')^2}+\CK_{-1}^{<<}\sr _3^<+ \CK_{1}^{<>}\srb _1^>=0.
\la{sigmatwom}
\ee
The kernels in the first and the third term combine to
\be \CK_1^{<>} (u, u') + {2\over\pi}{1\over (u-u')^2} = {1\over \pi}
\, \p_{u}\, \ {\sinh s' \cosh s + {2\over \pi} (s-s') \sinh^2 s' \over
\cosh(s-s')}\; .  \la{Kelem}
\ee
The procedure to obtain the previous expression was the following: we
expressed the first two terms $\chi_0$ and $\chi_1$ in the dressing
phase in terms of the variables $p$ and $\beta'$ and expanded them up
to the desired order in $\e$.  Then we expressed the results in terms
of the variables $s,s'$ by using the relations
$$u=\cos p/2=\tanh s\;, \qquad u'=\cosh \beta'/2=\coth s'\;.$$
Integrating the product of the kernel (\ref{Kelem}) with $\srb_1^>= -
2 \cosh (s/2)$ gives zero, so we deduce that
\eqn\srbp{ \sr ^{_<}_3(u)=0\;.  \la{srbp} }

\noindent
{\bf $\bullet$ order $\e^{3}$}

\medskip
\noindent
At this order,  equation (\ref{eqin}) gives
$$
{2\over\pi} \pint_{-\infty}^\infty d u'\frac{\sr_2(u')}{ (u-u')^2} +
\CK_{-1} \sr_4 + \CK_0 \sr_3 + \CK_1\sr_2+\CK_2\srb_1=0\, .
$$

\medskip

\noindent The subleading density $\sigma^{_>}_2(u)$, which we need in
order to compute the subleading term in $f(g)$, is supposed to be
determined by the equation with $|u|>1$
\be \frac{2}{\pi}\pint_{>}
du'\;\frac{\sigma_2^>(u')}{(u-u')^2}+\CK_{1}^{>>}
\sigma_2^>=-\CK_{2}^{>>}\bar\sigma_1^>\;.
\ee
After integrating with respect to $u$ and passing to the $s$
parametrization, this equation takes the same form as (\ref{intsr}),
but with different inhomogeneous term 
\begin{equation}
{1\over \pi} \pint _{-\infty}^\infty ds' \sigma^{_>}_2(s') \sinh s' \,
\coth{s-s'\over 2}\, = - {1\over 2\pi \sinh s} \int_{-\infty}^\infty
ds' \srb_1^{_>}(s') \, \phi_2^{>>}(s,s') \, .  \la{inteqc}
\end{equation}
The inhomogeneous term in the r.h.s. can be computed from the
expression
% \begin{eqnarray} \theta_1^{>>}=(2\e)^2\frac 1{2\pi^2} \frac
% 1{(\cosh\frac{\b_1}{2}-\cosh\frac{\b_2}{2})^2}\frac 1{\sinh\frac
% {\b_1}{2}\sinh\frac {\b_2}{2}}\times\\ \times\left[\left(\cosh\frac
% {\b_1}{2}-\cosh\frac{\b_2}{2}\right)+\left(1-\cosh\frac{\b_1}{2} 
%\cosh\frac{\b_2}{2}\right)\log\frac
% {\tanh\frac{\b_1}{4}}{\tanh\frac{\b_2}{4}}\right].  \end{eqnarray}
%
\begin{equation}
\phi_2^{>>}(s,s')={4(\sinh s\sinh s')^2 \over \pi}\ {(s-s') \cosh
(s-s')-\sinh (s-s')\over \sinh^2(s-s')},
\end{equation}
which, after being integrated with $\srb_1^> (s) =-\frac{ 2}{\pi}
\cosh(s/2) |u'(s)|$ gives
\begin{equation}
   - {1\over 2\pi \sinh s} \int_{-\infty}^\infty {ds' }\, \bar
   \sigma_1^>(s') \, \phi_2^{>>}(s,s') =-{2\over \pi} \sinh s
   \sinh(s/2)\, .
\end{equation}
After the redefinition (\ref{defor}) we again reduce the integral
equation to a Riemann boundary value problem 
\begin{equation}
     R(\o+i0)+R(\o-i0)= 2 \pint _0^{\infty} {d\o' } { r(\o')\over
     \o-\o'} = - { \sqrt{\o} (\o-1)(\o^2-1)\over 2 \o^2}\, .
\end{equation}
The solution is
$$
R(s)=- {s \sinh s \cosh(s/2)\over 2\pi}, \qquad r(s)= {s \sinh s \sinh
(s/2)\over 2\pi^2},
$$
which gives for the $\e^2$-correction to the density
\eqn\srtwo{ \sr_2^>(s) = {s \sinh (s/2)\over 2\pi^2}.  \la{srtwo} }
This correction to the density grows exponentially at $s\to\infty $
and is not normalizable\footnote{Strictly speaking, the maximum value
of $s$ to which we should integrate is $s_{{\rm max}}=\K/2\simeq -\ln
\e$.  If we stop the integral at $s_{{\rm max}}$, we will obtain a
contribution to the universal scaling function of the order $\ln
\e/\sqrt{\e}$, which violates the hypothesis that the expansion is in
integer power of $\e$.  Also, it signals that the limit $\e\to 0$ is
not uniform, and this is in accordance with the fact that in the
near-flat space regime all the terms $\chi_n$ contribute
\cite{Maldacena-Swanson}.}.  The significance of the result
(\ref{srtwo}) is that the higher order corrections to the density
become large in the vicinity of the points $u=\pm 1$, which correspond
to $s=\pm\infty$.  Therefore, in order to determine the next
correction to the scaling function $f(g)$ we have to take into account
the contribution coming from the points $u=\pm 1$ (the regimes $\pm$).
The equations starting with (\ref{sigmatwom}) should be corrected
accordingly.  We leave this problem for future work.

 \section{Outlook}
\label{out} \setcounter{equation}{0}

 We presented a method to treat the strong coupling limit of the Bethe
 ansatz equations, which can be used to extract the spectrum of
 anomalous dimensions.  We found it useful to express the spectral
 parameter appearing in the Bethe ansatz equations by means of a
 complexified momentum, with real part $p$ and imaginary part $\beta$
 \begin{equation}
 x^\pm=e^{\beta/2\pm i p/2}\, .
 \end{equation}
It is worth noticing that these variables already appeared in
\cite{RSS}, where the BDS magnons were interpreted as bound states of
more fundamental, fermionic excitations.  There, $p$ is the momentum
of the bound states, while $\beta$ is related to the ``size'' of the
bound state.
 
 Furthermore, the functions of $p$ and $\beta$ are naturally expressed
 in terms of elliptic functions.  Our elliptic parametrization is a
 version of that in \cite{Janik}.  In the strong coupling limit, the
 elliptic parametrization degenerates into a hyperbolic one.  In order
 to take into account properly the different regimes, we used
 different expansions of the elliptic functions in terms of hyperbolic
 functions, depending on the value of the the elliptic parameter $s$:
 around $s=0,\ \K$ and $\pm \K/2$.  These regimes correspond to the
 ``plane-wave'' \cite{BMN}, the ``giant magnons'' \cite{Maldacena,
 AZgiant} and the ``near-flat space'' regions, recently characterized
 in \cite{Maldacena-Swanson}, respectively.  It is interesting to note
 that the momentum $p$ and the energy $\varepsilon\equiv E/4g$ of the
 excitations have, in the plane wave and near-flat space regime,
 relativistic-like expressions
\begin{eqnarray}
&& p=\e \sinh s \qquad \varepsilon=\e (\cosh s-1)
\qquad {\rm (plane \ wave )}\\
 &&p=\pm 2\sqrt{\e}\;e^{\pm s}\;\qquad \varepsilon=2\sqrt{\e}\; e^{\pm
 s} \qquad \ {\rm (near\ flat\ space)}
 \end{eqnarray}
 with $s$ playing the role of the rapidity variable.  However, only
 the positive branch of the energy, corresponding to particles, appear
 in the sectors we consider, sectors which should be closed at any
 order in perturbation theory.  It is important to understand how the
 antiparticle branch will appear; most probably by another copy of the
 sector which joins in.  An example of how this should happen is
 offered by the $su(2)$ principal chiral model \cite{joem, Volodya},
 where the particles and antiparticles correspond to the two copies of
 $su(2)$-symmetric excitations.
 
 The strong coupling limit of the kernels without the dressing factor
 is relatively simple.  The giant magnons interact through a delta
 term, typical for the statistical repulsion.  In the $su(1|1)$
 sector, there is also a ``length-changing'' term.  The excitations of
 the plane-wave type do not interact with each other, at leading
 order, but they interact with the giant magnons.  The excitations in
 the near-flat space regime do not contribute to the leading term.
 
The strong coupling limit of the dressing kernel is more involved.
For the giant magnons and plane-wave regions, we can do it safely with
the strong coupling expansion of the dressing kernel \cite{AFS, HL,
BHL}, at least for the first few orders in $\e$.  For the near-flat
space limit, however, as it was pointed out in
\cite{Maldacena-Swanson}, all the terms in the strong coupling series
contribute to the leading order.    
 
 When we consider the kernels with the dressing phase, the repulsion
 of the giant magnons is so large that their density vanishes in the
 leading order.  For $su(1|1)$ and $su(2)$ sectors, most of the
 magnons are concentrated in the near-flat space regime.  Fortunately,
 although we do not have an explicit expression of the kernel in this
 regime, it is possible to obtain the energy for the highest excited
 state at leading order.
 
 The most exciting application of our method is to obtain the strong
 coupling expansion of the twist-two operator anomalous dimension and
 to compare with the string computations \cite{GKP,FT}.  Several
 recent analytical and numerical works \cite{KL06, Benna-I, Benna2}
 were devoted to this task.  In particular, \cite{Benna2} obtained the
 leading term, using the Fourier transform representation.  They also
 tried to obtain the result by a method very similar to ours, and
 found that the first two orders of the equation are not sufficient to
 fix the leading order.  Here, we show that it is possible to obtained
 the leading order, by going an order higher.  We also show that, in
 order to compute the higher correction, we have to take into account
 the contribution from the near-flat space regime.
  
  \medskip \noindent {\bf Note added.} An integral representation of
  the dressing phase, seemingly related to (\ref{drekint}), was used
  in \cite{Dorey:2007xn} to explore the analytic properties of the
  $S$-matrix .
  %
   
%%%%%%%%%%%%%%%%%%%%%%%%%%%%%%% 

\bigskip
\leftline{\bf Acknowledgments}

\noindent D.S.~and I.K.~thank the {\it Albert Einstein Institute},
Potsdam, and {\it Kavli Institute for Theoretical Physics}, Santa
Barbara, where part of this work was done, for hospitality.  D.S.
thanks Matthias Staudacher for collaboration in early stages of this
project.  We thank Gleb Arutyunov, Matteo Beccaria, Marcus Benna, Nick
Dorey, Burkhard Eden, Sergey Frolov, Igor Klebanov, Lev Lipatov,
Matthias Staudacher and Arkady Tseytlin for helpful discussions and
Stefan Zieme for spotting many typos in the first version of the
paper.  This work has been partially supported by the National Science
Foundation under Grant No.  PHY99-07949, by the European Union through
ENRAGE network (contract MRTN-CT-2004-005616) and the by ANR programs
GIMP (contract ANR-05-BLAN-0029-01) and INT-AdS/CFT (contract
ANR36ADSCSTZ)

 %%%%%%%%%%%%%%%%%%%%%%%%%%%% 

\appendix

  \section{ The elliptic parametrization }
 \setcounter{equation}{0}

 The  modulus of the elliptic map is
 \eqal\nome{ k= {1\over \sqrt{1+\e^2}}, \qquad k' = {\e \over
 \sqrt{1+\e^2}}.  }
 The real and imaginary parts of the complexified momentum 
\eqn\pofubis{ 
p(\s) =2 k'  \int _0^s { dv\over \dn v}
%= \pi - 2\, {\rm am} (K-\s, k)
,\quad  \b(\s)=  \ln{1+k'\over 1-k'} -2  \int _K^s \cs v\, dv 
%=  -i\pi +2i {\rm am}  (iK' - \s, k).
}
have the symmetries
 \eqal\sympb{ p(2K-u) &= &2\pi - p(u), \qquad \b(2K- u)= \b(u) \cr
 p(u) &=&-p(-u), \quad\qquad\b(-u)=\b(u)\mp2i\pi\, .
% \b( 2i K'-u)= \b( u)
} The phase in the Bethe equations depends on $\b$ and $p$ through
   \eqal\pebeta{ && \cosh\half\b ={1 \over k} \, \ns \s, \quad \sinh {
   \half \b } ={1 \over k }\, \ds \s ,\cr && \cos \half p \ \ =
   \cd\s,\qquad \sin \half p \ = k' \, \sd \s
%\cr &&\cr && \p_\s p = {2k' \nd \s} , \qquad \ \p_\s \b = - 2 \, \cs
%s
.
 }
 \eqal\qofu{ e^ {\pm \b /2} &=& {1\pm \dn \s \over k \, \sn \s}
%={\dn \st \pm k'\over k\cn \st}
, \quad e^{\pm ip/2} = \cd\s \pm i k' \, \sd \s \cr &&\cr \p_\s p &=&
{2k' \nd \s} , \qquad \ \p_\s \b \ = - 2 \, \cs s .  
\la{qofuA}}
 Below we give asymptotic expressions of these functions in
the limit $k'\simeq \e \to 0$, where
\eqal\hyperb{ K \approx \ln {4\over \e},\quad K' \approx{\pi\over 2},
\quad \tilde q = e^{-\pi K/K'}\approx {\e^2\over 16}.  }
In this limit it is appropriate to use the expansions of the three
elliptic functions in terms of hyperbolic functions of $v= {\pi s/2
K'}$:
\eqal\snDual{ \sn \s &=&{\pi\over 2 k K'}\left( \tanh v +
4\sum_{n=1}^{\infty} (-)^{n} {\tilde q^{2n}\over 1+\tilde q^{2n}}
\sinh 2nv \right) 
\cr 
\cn \s &=&{\pi\over 2 k K'}\left( {1\over \cosh
v} + 4\sum_{n=1}^{\infty} (-)^{n} {\tilde q^{2n-1}\over 1+\tilde
q^{2n-1}} \cosh (2n-1)v \right) 
\cr
\dn \s &=&{\pi\over 2 K'}\left(
{1\over \cosh v} - 4\sum_{n=1}^{\infty} (-)^{n} {\tilde q^{2n-1}\over
1-\tilde q^{2n-1}} \cosh (2n-1)v \right) 
.}
In the limit $\e\to 0$ and up to $o(\e^4)$ terms the three functions
are given by \cite{B-F}
\eqal\trigSN{ \sn \s &\simeq & \tanh \s\ + \ \e^2 \, { (\sinh \s \cosh
\s - \s) \over 4 \cosh ^2\s } 
\cr 
&&\cr
\cn \s &\simeq & {1 \over
\cosh \s } \ -\ \e^2\, \frac{ \tanh \s}{4} \(\sinh \s - {\s\over \cosh
\s}\) \cr &&\cr \dn \s &\simeq & {1\over \cosh \s } \ +\ \e^2\, \frac{
\tanh \s}{4} \(\sinh \s + {\s\over \cosh \s}\) .
  \label{asymh}
  }
We used only the leading term; it gives a good approximation for $s\in
I^{_>} = [-\hf K, \hf K]$.  Similarly, for the interval $\s\in I^{_<}
$ we can use the approximation
   \eqal\trigSN{ \sn (\K-\s) \ \simeq \ \ \ \cd \s &\simeq& 1\cr \cn
   (\K-\s) \ \simeq \ k' \sd \s &\simeq& \e \sinh \s \ \cr \dn (\K-\s)
   \ \simeq\ k'\nd \s &\simeq& \e\cosh\s\, .  }

  \bigskip
  
For the functions that enter in the definition of the kernel we find:
\bigskip

  (a) \    In the interval $  I^{_>} $: 
   \eqal\bminus{ \cosh(\b/2)= \coth \s,\ \ \sinh (\b /2 )= {1\over
   \sinh \s} ,\ \ \ \b' = -{2 \over\sinh \s} , &\no\\
&\no \\
    \cos(p/2)=1, \ \ \sin( p/2)= k' \sinh \s, \ \ \ \p_\s p = 2 k'
    \cosh \s\cr &\no\\
     u'= -{1\over\sinh^2 \s} ,\quad
     u  =  \coth \s
   \cr
  &\no\\
  e^{\b/2}= \coth{s\over 2}, \qquad e^s=\sqrt{u+1\over u-1}
  \label{triga}}

  (b) \ In the interval $ I^{_<}$, after replacing $ s \to K-\s$:
     \eqal\pbus{ \cosh (\b/2)=1,\ \ \ \sin (\b/2)= k' \cosh \s, \ \ \
     \p_\s \b = -2 k' \sinh \s\, , &\no\\
   &\no\\
   \cos(p/2)= \tanh \s,\ \ \ \sin(p/2)={1/\cosh \s}, \ \ \ \p_\s p =
   {2/ \cosh \s}\cr &\no\\
 u'= {1\over \cosh^2\s} \, , \quad u = \tanh\s, &
 \no\\
   &\no\\
    e^{ip/2}= {1+i e^{-\s}\over 1-i e^{-\s}}\, , \qquad
    e^{\s}=\sqrt{1-u\over 1+u} \la{trigb} }
    %
%     The quantities in the two intervals are related by $$ \b(\s+i
%     {\pi\over 2}) = i p (\st ).  $$
%     
 
\bigskip
\noindent

  \bigskip

\noindent Plugging these expressions in (\ref{Kssone}) we evaluate the
kernel in the four possible regimes.  In the sector $\s, \s_1 \in
I^{_>} $ the numerator in (\ref{Kssone}) is of order $\e$, while the
denominator remains finite.  Therefore \eqn\Kstst{ \Kpp(\s,\s_1)=0.  }
For the non-diagonal elements of (\ref{ppmatrix}) we obtain
\eqal\tKpmmp{ \Kpm(\s ,\s _1) = K(\s, \K-\s _1 ) && \simeq {1\over 4
\pi }\, { \b ' \, \sin( {p_1/2) } \over \cosh ( \b /2) - \cos (p _1/2)
} 
\cr && \cr && \cr 
&& \simeq {1\over 2 \pi } \ {1\over \cosh(\s - \s
_1)}\, , 
\cr && \cr && \cr
\Kmp(\s ,\s_1) = K(\K-\s , \s_1 ) && \simeq
\frac{1}{4\pi}\, p' - {1\over 4 \pi }\, {p' \, \sinh (\b _1/2) \over
\cosh( \b_1/2) - \cos (p/2 ) }
\cr && \cr &&
\simeq \frac{1}{2\pi }\,
\frac{1}{\cosh \s }- \frac{1}{2\pi }\, \frac{1}{\cosh (\s -s_1)} 
.  }
Finally, if both arguments are in $ I^{_<} $, then the kernel $
\Kmm(\s, \s_1)= K(\K-\s, \K-\s _1 ) $ vanishes except near the double
pole of the denominator at $\s=\s_1$, where it can be approximated by
a delta-function:
 \eqal\Kupvp{
 \Kmm(\s ,\s _1)=K(\K-\s, \K-\s _1 ) 
 && \simeq
 \frac{1}{4\pi}\, p' - {1\over 2 \pi }\, { p' \, \b \, \over \b ^2 +
 \sin^2 (p-p_1)/2} 
 \cr &&\cr
 && \simeq \frac{1}{4\pi}\, p'- |\p_sp|
 \delta \, ( p-p_1) 
 \cr &&\cr &&
 \simeq \frac{1}{2\pi}\, \frac{1}{\cosh
 s}-\delta(\s -\s _1).  }

 \bigskip
\bigskip

$\bullet$ {\it The regime $\s \simeq \pm  K/2$}

\bigskip

  % The points $s=\pm K/2$, which are the fixed points of the
  % transformation $s\to K-s$, correspond to the points
  % $u=\pm\sqrt{1+\e^2}$.  In the strong coupling limit, $\e \to 0$,
  % these are the points $u=\pm1$, and they correspond to meeting
  % points $s, \tilde s \to\pm\infty$ of the two parametrizations
  % (\ref{ugup}).  These points will be very important in the
  % following, and we need a better parametrization for them.

Assume that $\s =\pm K/2 +\y$, where $\y\ll \sqrt{\e}$.  We first
evaluate the three basic Jacobi elliptic functions for the shifted
argument: 
\eqal\halfK{ {\sn(s\pm \half \K) \over \sn\half \K}&=& {k'
\sd \pm \cn \, \over \cn^2 + k' \sn^2 }(\y)
% =\pm \ {1\mp k' {\rm sc}\, \y \, \nd \y\over 1+k' {\rm sc}^2\y}\ \dc
% \y
   \cr\cr {\cn(s\pm \half \K) \over \cn\half \K} & =& {\cn \mp
   \sn\dn\over \cn^2 + k'\sn^2}(\y)
%  = {1\mp \dc \y \, \sn \y \over 1+k' {\rm sc}^2\y}\ \nc \y
\cr\cr {\dn(s\pm \half \K) \over \dn\half \K} &=& { \dn \mp (1-k')
\sn\cn\over \cn^2 +k'\sn^2}(\y) \,.
}
When $k'\approx \e\to 0$, the values of the three functions at $s=\hf
\K$ are 
\eqal\haK{ &&\sn( \half \K) = \ {1 / \sqrt{1+k'}} \ \ \
\approx 1-\half \e,\cr && \cn( \half \K) = \sqrt{k' / (1+k')}\approx
\sqrt{\e},\cr && \dn( \half \K) =\qquad \sqrt{k' }\quad \ \ \ \approx
\sqrt{\e}, }
so that  
%$\dn(\pm \half \K)  \sn(\pm \half \K) 
%= \pm  \cn(\pm \half \K)  .
% $
%
  \eqal\halfK{
 \sn(\pm \half \K +\y)&=&
 \pm(1-\half  \e\,  e^{\mp 2 \y})
% = 
%   \pm\  (1-\half \e)
%  \(1  \pm  \e\,  \sinh\y \  e^{\mp s} \)
  \no
 \\
\cn(\pm \half \K +\y)&=& \sqrt{\e}\ e^{\mp \y} \( 1- \e \sinh^2 \y\)
\no \\
 \dn(\pm \half \K +\y)&=& \sqrt{\e}\ e^{\mp \y} \(1\pm \e \sinh \y\cosh\y\)
 }
 From here one finds for the asymptotics of the functions $p(\s)$,
 $\b(\s)$ and $u(\s)$ \\
 for $\s = \pm \K/2+\y $:
 \eqal\uKtwo{
 %\cosh [\half\b(\pm \half \K + \y) ] &=& 1+ \half \e\, e^{\mp 2
 %\y}\cr &&\cr
\sinh(\b/2)
%= \sinh {\b(\pm \half \K + \y) \over 2}
 &=& \pm \sqrt{\e}\, e^{\mp \y} \( 1 + \half \e\, \cosh 2\y\)\cr &&\cr
 \sin (p /2)
%= \sin {p(\pm \half \K + \y) \over 2}
 &=& \pm \sqrt{\e}\, e^{\pm \y} \( 1 - \half \e\, \cosh 2\y\)\cr &&\cr
 u
%  =u(\pm \half \K + \y)
 &=& \pm 1 - \e \sinh 2y + o(\e^2).  }

 \section{ Evaluation of  $K_d^0$}
   
   \setcounter{equation}{0}

Here we will evaluate the integral (\ref{Lnn}) for the strong 
coupling limit of $K_d$. We have 
\eqal\KpKm{ K_-(x_1,x) &=& \frac{K^p_- dp_1+ K^\b_- d\b_1}{4\pi} \cr 
K_+(x, x_2)&=& \frac{K_+^p dp + K_+^\b d\b}{2\pi} \cr {dz }\ \ \ & 
=& \half Z^pdp+\half Z^\b d\b, }
where
\eqal\Koefs{K_-^p&=& {\cos(p-p_1) - e^{-|\beta+\b_1|}\over H_1}, 
\quad K_-^\b =- {\sin(p_1-p) \over H_1}\, ; \cr K_+^p&=& 
{\sinh{\beta+\b_2\over 2}\cos{p- p_2\over 2}\over H_2}, \quad K_+^\b 
= - {\sin{p- p_2\over 2}\cosh{\beta+\b_2\over 2} \over H_2}\, ;\cr 
\no
\\
 Z^p&=&   \sinh(\b/2)\cos(p/2), \qquad  Z^\b= \sin(p/2)\cosh(\b/2) \, ; 
 \no \\
 \cr
 H_a &=& \cosh(\beta+\b_a)-\cos(p- p_a),\quad a=1,2\,
\la{Coefs}.
 }

\bigskip \noindent The integral to evaluate is
\eqal\FffK{\label{FffK} K_d^0(x_1,x_2) &=&-\frac{1}{2 \pi^2 \e} \int 
_0^{2\pi } dp\int_0^\infty d\b\ (K_-^p dp_1 + K_-^\b d\b_1)
 (Z^p K_+^\b -Z^\b
    K^p_ +   )\cr
 &=& \frac{1}{4 \pi^2 \e}   
\int _0^{2\pi } dp\int_0^\infty d\b\ [ A(p, \b)d p_1 + B(p,\b) 
d\b_1] \, , }
 where
\eqal\Aone{\la{Aone} A(\b,p) &=& ( \cos(p-p_1)-e^{-|\b+\b_1|}) \ 
G(\b,p),\cr
\no\\
B(\b,p) &=& - 
  \sin(p_1-p) \ G(\b,p),
\cr
\no\\
G(\b,p)&=& {\sin(p- {p_2\over 2})\sinh( \b+{\b_2\over 2}) + 
\sinh{\b_2\over 2} \sin{p_2\over 2} \over \(\cosh(\beta+\b_1)-\cos( 
p_1-p)\)\( \cosh(\beta+\b_2)-\cos(p- p_2)\) }\, . }
% 

%Integrands of (\ref{FffK}) are the rational expressions of 
%trigonometric and hyperbolic functions and thus the integral can be 
%calculated explicitly. 

The strategy of calculation is the following: first we take an 
integral over $p$, which  is of the form: 
\begin{equation} 
    I=\int\limits_0^{2\pi} dp\, R (\cos p,\sin p), 
\end{equation} 
where $R$ is a rational function. By symmetrization $p\rightarrow 
-p$ we reduce the integral to 
\begin{equation} 
    I=\int\limits_0^{2\pi} dp\, \tilde R (\cos p,\sin^2 p)
\end{equation}
and, after   substituting $\cos p=-t$, it can be evaluated  by 
taking the residues.

The integral over $\beta$ becomes, after the substitution $\beta=-\log
x$, an integral from 0 to 1 of a rational expression and can also be
performed explicitly.  As a final result, we obtain
\begin{equation} 
     \int\limits_0^\infty d\beta \int\limits_0^{2\pi} dp\, A(p,\beta)=
     i\pi\(\frac 1{x_2^-}+\frac 12(x_1^+-\frac 1{x_1^+})\log \frac
     {1-\frac 1{x_1^+x_2^-}}{1+\frac 1{x_1^+x_2^-}}-{\rm c.c.}\),
\end{equation} 
\begin{equation}
 \int\limits_0^\infty d\beta \int\limits_0^{2\pi} dp\, B(p,\beta)=
 \pi\(\frac 1{x_2^-}+\frac 12(x_1^+-\frac 1{x_1^+})\log \frac {1-\frac
 1{x_1^+x_2^-}}{1+\frac 1{x_1^+x_2^-}}+{\rm c.c.}\).
\end{equation}
From here we deduce the expression for 
 $K_d^0$,  presented in  (\ref{kdo}).


\bigskip
\bigskip


\begin{thebibliography}{10} 
 \raggedright
 \small 
 \parskip 0pt




%1% 
\bibitem{MZ02}
J.~A.~Minahan and K.~Zarembo, \textit{``The Bethe-ansatz for
{$\mathcal{N}=\mathord{}$4} super Yang-Mills''},
\textsf{JHEP~0303,~013~(2003)}, \texttt{{hep-th/0212208}}.
%
%2
\bibitem{BS03}
N.~Beisert and M.~Staudacher, \textit{``The
{$\mathcal{N}=\mathord{}$4} SYM Integrable Super Spin Chain''},
\textsf{Nucl.~Phys.~B670,~439~(2003)},%
\texttt{{hep-th/0307042}}.
%
%3
\bibitem{BKS}
N.~Beisert, C.~Kristjansen and M.~Staudacher, \textit{``The Dilatation
Operator of {$\mathcal{N}=\mathord{}$4} Conformal Super Yang-Mills
Theory''}, \textsf{{Nucl.~Phys.~B664,~131~(2003)}},
\texttt{{hep-th/0303060}}.
%
%4
\bibitem{MaldaAdS}
J.~M.~Maldacena, \textit{``The large N limit of superconformal field
theories and supergravity''},
\textsf{Adv.~Theor.~Math.~Phys.~2,~231~(1998)},
\texttt{{hep-th/9711200}}.
%
%5
\bibitem{GKP98}
S.~S.~Gubser, I.~R.~Klebanov and A.~M.~Polyakov, \textit{``Gauge
theory correlators from non-critical string theory''},
\textsf{Phys.~Lett.~B428,~105~(1998)}, \texttt{{hep-th/9802109}}.
%
%6
\bibitem{Witten98}
E.~Witten, \textit{``Anti-de Sitter space and holography''},
\textsf{Adv.~Theor.~Math.~Phys.~2,~253~(1998)},
\texttt{{hep-th/9802150}}.
%
%
%7
\bibitem{StaudS}
  M.~Staudacher, \textit{``The factorized S-matrix of CFT/AdS,''}
  \textsf{JHEP {\bf 0505} (2005) 054} \texttt{hep-th/0412188}.
  %
  %8
\bibitem{BSansaetze}
  N.~Beisert and M.~Staudacher, \textit{``Long-range PSU(2,2|4) Bethe
  ansaetze for gauge theory and strings'',} \textsf{Nucl.\ Phys.\ B
  {\bf 727}, 1 (2005)}, \texttt{hep-th/0504190.}
%9
\bibitem{Beisert05}
  N.~Beisert, \textit{``The Analytic Bethe Ansatz for a Chain with
  Centrally Extended su(2|2) Symmetry'',} \textsf{J.\ Stat.\ Mech.\
  {\bf 0701}, P017 (2007)}, \texttt{nlin.si/0610017.}
   %10
\bibitem{Janik}
  R.~A.~Janik, \textit{``The $AdS(5) \times S^5$ superstring
  worldsheet S-matrix and crossing symmetry'',} \textsf{Phys.\ Rev.\ D
  {\bf 73}, 086006 (2006)}, \texttt{hep-th/0603038.}
  %
  %11
\bibitem{AFS}
G.~Arutyunov, S.~Frolov and M.~Staudacher, \textit{``Bethe ansatz for
quantum strings''}, \textsf{{JHEP~0410,~016~(2004)}},
\texttt{{hep-th/0406256}}.
%
%12
\bibitem{HL}
R.~Hern{\'a}ndez and E.~L{\'o}pez, \textit{``Quantum corrections to
the string Bethe ansatz''}, \textsf{{JHEP~0607,~004~(2006)}},
\texttt{{hep-th/0603204}}.
%
%13
\bibitem{BHL}
N.~Beisert, R.~Hern\'andez and E.~L\'opez, \textit{``A
Crossing-Symmetric Phase for $AdS_5 \times S^5$ Strings''},
\texttt{{hep-th/0609044}}.
%14
\bibitem{BES}
  N.~Beisert, B.~Eden and M.~Staudacher, \textit{``Transcendentality
  and crossing'',} \textsf{J.\ Stat.\ Mech.\ {\bf 0701}, P021 (2007)},
  \texttt{hep-th/0610251.}
  %15
  %
\bibitem{KLOV}
A.~V.~Kotikov, L.~N.~Lipatov, A.~I.~Onishchenko and V.~N.~Velizhanin,
\textit{``Three-loop universal anomalous dimension of the Wilson
operators in {$\mathcal{N}=\mathord{}$4} SUSY Yang-Mills model''},
\textsf{Phys.~Lett.~B595,~521~(2004)}, \texttt{{hep-th/0404092}}.
%
%16
\bibitem{Moch}
S.~Moch, J.~A.~M.~Vermaseren and A.~Vogt, \textit{``The three-loop
splitting functions in QCD: The non-singlet case''},
\textsf{Nucl.~Phys.~B688,~101~(2004)}, \texttt{{hep-ph/0403192}}.
%
%17
\bibitem{Bern03}
C.~Anastasiou, Z.~Bern, L.~J.~Dixon and D.~A.~Kosower,
\textit{``Planar amplitudes in maximally supersymmetric Yang-Mills
theory''}, \textsf{Phys.~Rev.~Lett.~91,~251602~(2003)},
\texttt{{hep-th/0309040}}.
%
%18
\bibitem{Bern05}
Z.~Bern, L.~J.~Dixon and V.~A.~Smirnov, \textit{``Iteration of planar
amplitudes in maximally supersymmetric Yang-Mills theory at three
loops and beyond''}, \textsf{Phys.~Rev.~D72,~085001~(2005)},
\texttt{{hep-th/0505205}}.
%
%19
\bibitem{Bern06}
Z.~Bern, M.~Czakon, L.~J.~Dixon, D.~A.~Kosower and V.~A.~Smirnov,
\textit{``The Four-Loop Planar Amplitude and Cusp Anomalous Dimension
in Maximally Supersymmetric Yang-Mills Theory''},
\texttt{{hep-th/0610248}}.
%
%20
\bibitem{CSV}
  F.~Cachazo, M.~Spradlin and A.~Volovich, \textit{``Four-loop cusp
  anomalous dimension from obstructions,''} \texttt{hep-th/0612309.}
  %21
%  
\bibitem{GKP}
  S.~S.~Gubser, I.~R.~Klebanov and A.~M.~Polyakov, \textit{``A
  semi-classical limit of the gauge/string correspondence'',}
  \textsf{Nucl.\ Phys.\ B {\bf 636}, 99 (2002)},
  \texttt{hep-th/0204051.}
     %22
\bibitem{FT}
S.~Frolov and A.~A.~Tseytlin, \textit{``Semiclassical quantization of
rotating superstring in {$AdS_5 \times S^5$}''},
\textsf{JHEP~0206,~007~(2002)}, \texttt{{hep-th/0204226}}.
%23
\bibitem{KL06}
 A.~V.~Kotikov and L.~N.~Lipatov, \textit{``On the highest
 transcendentality in N = 4 SUSY'',} \texttt{hep-th/0611204.}
  % 
%24
\bibitem{Benna-I}
  M.~K.~Benna, S.~Benvenuti, I.~R.~Klebanov and A.~Scardicchio,
  \textit{``A test of the AdS/CFT correspondence using high-spin
  operators'',} \texttt{hep-th/0611135.}
    %25
    \bibitem{Benna2}
  L.~F.~Alday, G.~Arutyunov, M.~K.~Benna, B.~Eden and I.~R.~Klebanov,
  \textit{``On the strong coupling scaling dimension of high spin
  operators'',} \texttt{hep-th/0702028.}
  %26
\bibitem{RSZ}
  A.~Rej, M.~Staudacher and S.~Zieme, \textit{``Nesting and
  dressing,''} \texttt{hep-th/0702151.}
  %27
\bibitem{Gomez}
  C.~Gomez and R.~Hern\'andez, \textit{ ``Quantum deformed magnon
  kinematics,''} \texttt{hep-th/0701200}.
   %28
   \bibitem{Maldacena-Swanson}
  J.~Maldacena and I.~Swanson, \textit{``Connecting giant magnons to
  the pp-wave: An interpolating limit of $AdS(5) \times S^5$'',}
  \texttt{hep-th/0612079.}
  %29
 \bibitem{ES}
  B.~Eden and M.~Staudacher, \textit{``Integrability and
  transcendentality'',} \textsf{J.\ Stat.\ Mech.\ {\bf 0611}, P014
  (2006)}, \texttt{hep-th/0603157.}
  %30
    %
  \bibitem{BDS}
N.~Beisert, V.~Dippel and M.~Staudacher, \textit{``A Novel Long Range
Spin Chain and Planar {$\mathcal{N}=\mathord{}$4} Super Yang-Mills''},
\textsf{{JHEP~0407,~075~(2004)}}, \texttt{{hep-th/0405001}}.
%31
  %
  \bibitem{Korchemsky}
  G.~P.~Korchemsky, \textit{``Quasiclassical QCD pomeron'',}
  \textsf{Nucl.\ Phys.\ B {\bf 462}, 333 (1996)},
  \texttt{hep-th/9508025.}
  % 
%32
\bibitem{BMN}
D.~Berenstein, J.~M.~Maldacena and H.~Nastase, \textit{``Strings in
flat space and pp waves from {$\mathcal{N}=\mathord{}$4} {Super} {Yang
Mills}''}, \textsf{JHEP~0204,~013~(2002)}, \texttt{{hep-th/0202021}}.
%33
\bibitem{Maldacena}
  D.~M.~Hofman and J.~M.~Maldacena, \textit{``Giant magnons'',}
  \textsf{J.\ Phys.\ A {\bf 39}, 13095 (2006)},
  \texttt{hep-th/0604135.}
 %34
\bibitem{AZgiant}
  G.~Arutyunov, S.~Frolov and M.~Zamaklar, \textit{``Finite-size
  effects from giant magnons,''} \texttt{hep-th/0606126.}
  %35
\bibitem{Dorey}
  N.~Dorey, \textit{ ``Magnon bound states and the AdS/CFT
  correspondence''}, \textsf{J.\ Phys.\ A {\bf 39} (2006) 13119}
  \texttt{hep-th/0604175.}
  %36
  \bibitem{Chen:2006gq}
  H.~Y.~Chen, N.~Dorey and K.~Okamura, \textit{``On the scattering of
  magnon boundstates,''} \textsf{ JHEP {\bf 0611}, 035 (2006)}
  \texttt{:hep-th/0608047}.
  %37
%
  \bibitem{RSS}
  A.~Rej, D.~Serban and M.~Staudacher, \textit{``Planar N = 4 gauge
  theory and the Hubbard model'',} \textsf{JHEP {\bf 0603}, 018
  (2006)}, \texttt{hep-th/0512077.}
  % 38
%
\bibitem{Zarembo05}
  K.~Zarembo, \textit{``Antiferromagnetic operators in N = 4
  supersymmetric Yang-Mills theory'',} \textsf{Phys.\ Lett.\ B {\bf
  634}, 552 (2006)}, \texttt{hep-th/0512079.}
  %39
 \bibitem{Beccaria-I}
   M.~Beccaria and L.~Del Debbio, \textit{``Bethe Ansatz solutions for
   highest states in N = 4 SYM and AdS/CFT duality'',} \textsf{JHEP
   {\bf 0609}, 025 (2006)}, \texttt{hep-th/0607236.}
  %40
 \bibitem{Beccaria-II}
   M.~Beccaria, G.~F.~De Angelis, L.~Del Debbio and M.~Picariello,
   \textit{``Highest states in light-cone $AdS(5) \times S^5 $
   superstring'',} \texttt{hep-th/0701167.}
  % 
  %
   %41
\bibitem{SS05}
 D.~Serban and M.~Staudacher, unpublished, 2005
%42
\bibitem{faddeev}
  L.~D.~Faddeev, \textit{``How Algebraic Bethe Ansatz works for
  integrable model'',} \texttt{hep-th/9605187.}
  %43
\bibitem{feverati}
  G.~Feverati, D.~Fioravanti, P.~Grinza and M.~Rossi, \textit{``On the
  finite size corrections of anti-ferromagnetic anomalous dimensions
  in N = 4 SYM,''} \textsf{JHEP {\bf 0605} (2006) 068}
  \texttt{hep-th/0602189.}
  %44
  %
 \bibitem{ArutyunovTseytlin}
 G.~Arutyunov and A.~A.~Tseytlin, \textit{``On highest-energy state in
 the su(1|1) sector of N = 4 super Yang-Mills theory'',} \textsf{JHEP
 {\bf 0605}, 033 (2006)}, \texttt{hep-th/0603113.}
  %45
%
 \bibitem{B-F}
P. Byrds and M. Friedman, {\it Handbook of Elliptic Integrals for
Engineers and Physicists}, Springer-Verlag, Berlin 1954, p.25.
%46
\bibitem{LW}
E.~H.~ Lieb and F.~Y.~ Wu, \textit{``Absence of Mott transition in an
exact solution of the short-range, one-band model in one dimension,''}
\textsf{ Phys.  Rev.  Lett.  20 (1968) 1445; Erratum, ibid.  21 (1968)
192.  }
%47
\bibitem{Haldane}
  F.~D.~M.~Haldane, \textit{``Exact Jastrow-Gutzwiller Resonating
  Valence Bond Ground State Of The Spin 1/2 Antiferromagnetic
  Heisenberg Chain With $1/R^2$ Exchange,''} \textsf{ Phys.\ Rev.\
  Lett.\ {\bf 60}, 635 (1988).}
  %48
  \bibitem{Shastry}
   B.~ S.~ Shastry, \textit{``Exact solution of an S=1/2 Heisenberg
   antiferromagnetic chain with long-ranged interactions''}, \textsf{
   Phys.  Rev.  Lett.  60, 639 (1988).}
%
%49
\bibitem{Sutherland}
  B.~Sutherland, \textit{``Exact results for a quantum many body
  problem in one-dimension,''} \textsf{ Phys.\ Rev.\ A {\bf 4}, 2019
  (1971).}
  %50
  \bibitem{RuijSch}
  S.~N.~M.~Ruijsenaars and H.~Schneider, \textit{ ``A New Class Of
  Integrable Systems And Its Relation To Solitons,''} \textsf{ Annals
  Phys.\ {\bf 170} (1986) 370.}
 %51
 \bibitem{Ino}
  V.~ I.~ Inozemtsev, \textit{``Integrable Heisenberg-van Vleck chains
  with variable range ex- change,'' } \textsf{Phys.  Part.  Nucl.  34
  (2003) 166 [Fiz.  Elem.  Chast.  Atom.  Yadra 34 (2003) 332],}
  \texttt{hep-th/0201001}.
  %52
\bibitem{SS04}
D.~Serban and M.~Staudacher, \textit{``Planar
{$\mathcal{N}=\mathord{}$4} gauge theory and the Inozemtsev long range
spin chain''}, \textsf{{JHEP~0406,~001~(2004)}},
\texttt{{hep-th/0401057}}.
%53
\bibitem{LipatovPotsdam}
L.~N.~Lipatov, \textit{``Transcendentality and Eden-Staudacher
equation''}, Talk at Workshop on Integrability in Gauge and String
Theory, AEI, Potsdam, Germany, July 24-28, 2006,
{\texttt{http://int06.aei.mpg.de/presentations/lipatov.pdf}}.
%54
\bibitem{AF}
G.~Arutyunov and S.~Frolov, \textit{``On $AdS_5\times S^5$ string
S-matrix''}, \textsf{Phys.~Lett.~B639,~378~(2006)},
\texttt{{hep-th/0604043}}.
%
  %55
\bibitem{LCBA}
G.~Arutyunov, S.~Frolov, J.~Plefka and M.~Zamaklar, \textit{``The
Off-shell Symmetry Algebra of the Light-cone $AdS_5\times S^5$
Superstring''}, \texttt{{hep-th/0609157}}.
%56
\bibitem{AFsu11}
  G.~Arutyunov and S.~Frolov, \textit{`Uniform light-cone gauge for
  strings in AdS(5) x S**5: Solving su(1|1) sector,''} \textsf{ JHEP
  {\bf 0601} (2006) 055} \texttt{hep-th/0510208}.
  % 57
\bibitem{joem}
  J.~A.~Minahan, \textit{``The SU(2) sector in AdS/CFT,''}
  \textsf{Fortsch.\ Phys.\ {\bf 53} (2005) 828}
  \texttt{hep-th/0503143.}
  % 58
  \bibitem{Volodya}
  N.~Gromov and V.~Kazakov, \textit{``Asymptotic Bethe ansatz from
  string sigma model on $S^3 \times R$,''} \texttt{hep-th/0605026.}
  %59
\bibitem{Dorey:2007xn}
  N.~Dorey, D.~M.~Hofman and J.~Maldacena, \textit{``On the
  singularities of the magnon S-matrix,''} \texttt{hep-th/0703104.}
  
\end{thebibliography}
 \end{document}